\documentclass[11pt]{article}
\usepackage{a4p,cite,epsfig}

\begin{document}

\newcommand{\mco}{\multicolumn}
\newcommand{\antibar}[1]{\ensuremath{\mathrm{#1\overline{#1}}}}
\newcommand{\Aqq}{\ensuremath{A_{\mathrm{FB}}^{\qq}}}
\newcommand{\Afbb}{\ensuremath{A_{\mathrm{FB}}^{\bb}}}
\newcommand{\Afbc}{\ensuremath{A_{\mathrm{FB}}^{\cc}}}
\newcommand{\asy}{\ensuremath{A_{\mathrm{FB}}}}
\newcommand{\Zzero}{\ensuremath{\mathrm{Z}}}
\newcommand{\ee}{\ensuremath{\mathrm{e}^+\mathrm{e}^-}}
\newcommand{\qq}{\antibar{q}}
\newcommand{\bb}{\antibar{b}}
\newcommand{\cc}{\antibar{c}}
\newcommand{\uu}{\antibar{u}}
\newcommand{\dd}{\antibar{d}}
\newcommand{\eeqq}{\ensuremath{\ee\rightarrow\qq}}
\newcommand{\eebb}{\ensuremath{\ee\rightarrow\bb}}
\newcommand{\eecc}{\ensuremath{\ee\rightarrow\cc}}
\newcommand{\abscosthe}{\ensuremath{|\cos\theta|}}
\newcommand{\costh}{\ensuremath{\cos\theta}}
\newcommand{\swsqeffl}{\ensuremath{\sin^2\theta_{\mathrm{eff}}^{\ell}}}
\newcommand{\thwefff}{\theta_{\mathrm{eff}}^{{\rm f}}}
\newcommand{\thweffb}{\theta_{\mathrm{eff}}^{{\rm b}}}
\newcommand{\ga} {g_{\mathrm{A}}}
\newcommand{\gv} {g_{\mathrm{V}}}
\newcommand{\gae}{g_{\mathrm{Ae}}}
\newcommand{\gve}{g_{\mathrm{Ve}}}
\newcommand{\gaf}{g_{\mathrm{Af}}}
\newcommand{\gvf}{g_{\mathrm{Vf}}}
\newcommand{\gaq}{g_{\mathrm{Aq}}}
\newcommand{\gvq}{g_{\mathrm{Vq}}}
\newcommand{\Bzero}{\ensuremath{{\rm B}^0}}
\newcommand{\Bzeros}{\ensuremath{\mathrm{ B_s^0}}}
\newcommand{\Bplus}{\ensuremath{{\rm B}^+}}
\newcommand{\Bzerob}{\ensuremath{{\overline{{\rm B}^0}}}}
\newcommand{\bob}{\ensuremath{\Bzero\Bzerob}}
\newcommand{\mix}{\ensuremath{\overline{\chi}}}
\newcommand{\oeta}{\ensuremath{\overline\eta}}
\newcommand{\Afbzq}{\ensuremath{A^{0,\,{\mathrm{q}}}_{\mathrm{FB}}}}
\newcommand{\Afbzb}{\ensuremath{A^{0,\,{\mathrm{b}}}_{\mathrm{FB}}}}
\newcommand{\Afbzc}{\ensuremath{A^{0,\,{\mathrm{c}}}_{\mathrm{FB}}}}
\newcommand{\cAe}{\ensuremath{{\cal A}_{\mathrm{e}}}}
\newcommand{\cAq}{\ensuremath{{\cal A}_{\mathrm{q}}}}
\newcommand{\cAb}{\ensuremath{{\cal A}_{\mathrm{b}}}}
\newcommand{\cAc}{\ensuremath{{\cal A}_{\mathrm{c}}}}
\newcommand{\etal}{\mbox{{\it et al.}}}
\newcommand{\PLB}[3] {Phys.~Lett.\ {B#1} (#2) #3}
\newcommand{\PRL}[3] {Phys.~Rev.\ {Lett.~#1} (#2) #3}
\newcommand{\PRD}[3] {Phys.~Rev.\ {D#1} (#2) #3}
\newcommand{\PRP}[3] {Phys.~Rep.\ {#1} (#2) #3}
\newcommand{\NIM}[3] {Nucl.~Instr.\ {Meth.~#1} (#2) #3}
\newcommand{\NPB}[3] {Nucl.~Phys.\ {B#1} (#2) #3}
\newcommand{\CPC}[3] {Comp.~Phys.\ {Comm.~#1} (#2) #3}
\newcommand{\ZPC}[3] {Z.~Phys.\ {C#1} (#2) #3}
\newcommand{\JPH}[3] {J.~Phys.\ {#1} (#2) #3}
\newcommand{\JHP}[3] {JHEP\ {#1} (#2) #3}
\newcommand{\EPJ}[3] {Eur.~Phys.\ J.\ {C#1} (#2) #3}
\newcommand{\IJA}[3] {Int.~J.~Mod.~Phys.\ A.\ {C#1} (#2) #3}
\newcommand{\SJN}[3] {Sov.~J.~Nucl.~Phys.\ {#1} (#2) #3}
\newcommand{\cthr}{\ensuremath{\cos\theta_{\mathrm{T}}}}
\newcommand{\dEdx}{\ensuremath{\mathrm{d}E/\mathrm{d}x}}
\newcommand{\bl}{\ensuremath{\rm b \rightarrow \ell^- }}
\newcommand{\bD}{\ensuremath{\mathrm{b \rightarrow D} }}
\newcommand{\bu}{\ensuremath{\mathrm{b \rightarrow u} }}
\newcommand{\bcl}{\ensuremath{\mathrm{b \rightarrow c \rightarrow \ell^+}}}
\newcommand{\bcbarl}{\ensuremath{\mathrm{b \rightarrow \overline{c} \rightarrow \ell^-}}}
\newcommand{\btaul}{\ensuremath{\mathrm{b \rightarrow \tau^- \rightarrow \ell^-}}}
\newcommand{\bpsill}{\ensuremath{\mathrm{b \rightarrow J/\psi \rightarrow \ell^+\ell^-}}}
\newcommand{\cl}{\ensuremath{\mathrm{c} \rightarrow \ell^+ }}
\newcommand{\Rb}{\ensuremath{R_{\mathrm{b}}}}
\newcommand{\Rc}{\ensuremath{R_{\mathrm{c}}}}
\newcommand{\p}{\ensuremath{p}}
\newcommand{\pt}{\ensuremath{p_t}}
\newcommand{\jte}{\ensuremath{E\mathrm{^{vis}_{jet}}}}
\newcommand{\esub}{\ensuremath{E_{\mathrm{sub-jet}}}}
\newcommand{\jspt}{\ensuremath{(\sum \pt)_{\mathrm{jet}}}}
\newcommand{\sigl}{\ensuremath{ L/\sigma }}
\newcommand{\siglb}{\ensuremath{ (L/\sigma_L)_{1,2} }}
\newcommand{\sigb}{\ensuremath{ d/\sigma_d }}
\newcommand{\NETb}{\mbox{NETb}}
\newcommand{\NETc}{\mbox{NETc}}
\newcommand{\Dzero}{{\rm D}^0}
\newcommand{\Dplus}{{\rm D}^+}
\newcommand{\Dss}{\ensuremath{\mathrm{D^{**}}}}
\newcommand{\mean}[1]{\langle{#1}\rangle}
\newcommand{\meanxe}{\ensuremath{\mean{x_E}}}
\newcommand{\mxeb}{\ensuremath{\mean{x_E^{\mathrm{b}}}}}
\newcommand{\mxec}{\ensuremath{\mean{x_E^{\mathrm{c}}}}}
\newcommand{\sss}{\subsubsection*}
\newcommand{\nf}{\ensuremath{n_{\mathrm{F}}}}
\newcommand{\nb}{\ensuremath{n_{\mathrm{B}}}}
\newcommand{\ns}{\ensuremath{n_{\mathrm{S}}}}
\newcommand{\nt}{\ensuremath{n_{\mathrm{T}}}}
\newcommand{\pf}{\ensuremath{p_{\mathrm{F}}}}
\newcommand{\pb}{\ensuremath{p_{\mathrm{B}}}}
\newcommand{\ps}{\ensuremath{p_{\mathrm{S}}}}
\newcommand{\fraci}{\ensuremath{f_i}}
\newcommand{\fracij}{\ensuremath{f_{ij}}}
\newcommand{\hlf}{\mbox{$\frac 1 2$}}
\newcommand{\qrt}{\mbox{$\frac 1 4$}}

\begin{titlepage}
 
\begin{center}
{\Large \bf EUROPEAN ORGANIZATION FOR NUCLEAR RESEARCH}
\end{center}
 
\bigskip
 
\begin{flushright}
CERN-EP/2003-039 \\
11 July 2003 \\
\end{flushright}
 
\vspace{1cm}
 
\begin{center}
\Large{\bf  Measurement of  Heavy Quark Forward-Backward \\
Asymmetries and  Average B Mixing \\
Using Leptons in Hadronic Z Decays\\}
 
\vspace{1.3cm}
 \Large{\bf The OPAL Collaboration}

\vspace{10mm}

\end{center}
 
\vspace{0.1cm}
 
\begin{center}
{\large\bf Abstract\\}
\end{center}

A measurement of  the forward-backward asymmetries
of \eebb\ and \eecc\ events using  electrons and muons produced
in semileptonic decays of bottom and charm hadrons is presented.
The outputs of two neural networks designed to identify $\bl$
and $\cl$ decays
are used in a maximum likelihood fit to a sample of events containing
one or two identified leptons. The b and c quark forward-backward asymmetries
at three centre-of-mass energies $\sqrt{s}$
and the average B mixing parameter \mix\ are determined simultaneously in the 
fit. Using all data collected by OPAL near the \Zzero\ resonance,
the asymmetries are measured to be:
\begin{center}
\begin{tabular}{@{\Afbb = (}r@{.}l@{$\,\pm\,$}r@{.}l@{$\,\pm\,$}r@{.}l@{)\,\%\ \ \ 
                  \Afbc = (}r@{.}l@{$\,\pm\,$}r@{.}l@{$\,\pm\,$}r@{.}l@{)\,\%\ \ \ 
                  at $ \langle \protect\sqrt s \rangle = $ }r@{.}l@{}}
4&7 & 1&8 & 0&1 & $-$6&8 & 2&5 & 0&9 & 89&51 GeV, \\
9&72 & 0&42 & 0&15 & 5&68 & 0&54 & 0&39 & 91&25 GeV,\\
10&3 & 1&5 & 0&2 & 14&6 & 2&0 & 0&8 & 92&95 GeV.\\
\end{tabular}
\end{center}
For the average B mixing parameter, a value of:
$$
\mix = (13.12 \pm 0.49 \pm 0.42)\,\%
$$
is obtained. In each case the first uncertainty
is statistical and the second systematic.
These results are combined with other OPAL measurements of the b and c
forward-backward asymmetries, and used to derive a value for 
the effective electroweak mixing angle for leptons \swsqeffl\ of
$0.23238 \pm 0.00052$.

\vspace*{1.cm}
\vspace*{1.cm}
\begin{center}{\large
Submitted to Phys. Lett. B \\
}
\end{center}

\end{titlepage}
\begin{center}{\Large        The OPAL Collaboration
}\end{center}\smallskip
\begin{center}{
G.\thinspace Abbiendi$^{  2}$,
C.\thinspace Ainsley$^{  5}$,
P.F.\thinspace {\AA}kesson$^{  3}$,
G.\thinspace Alexander$^{ 22}$,
J.\thinspace Allison$^{ 16}$,
P.\thinspace Amaral$^{  9}$, 
G.\thinspace Anagnostou$^{  1}$,
K.J.\thinspace Anderson$^{  9}$,
S.\thinspace Arcelli$^{  2}$,
S.\thinspace Asai$^{ 23}$
% ...
}\end{center}
\medskip
$^{  1}$School of Physics and Astronomy, University of Birmingham,
Birmingham B15 2TT, UK
\newline
$^{  2}$Dipartimento di Fisica dell' Universit\`a di Bologna and INFN,
I-40126 Bologna, Italy
\newline
$^{  3}$Physikalisches Institut, Universit\"at Bonn,
D-53115 Bonn, Germany
\newline
% ...
$^{  5}$Cavendish Laboratory, Cambridge CB3 0HE, UK
\newline
% ...
$^{  9}$Enrico Fermi Institute and Department of Physics,
University of Chicago, Chicago IL 60637, USA
\newline
% ...
$^{ 16}$Department of Physics, Schuster Laboratory, The University,
Manchester M13 9PL, UK
\newline
% ...
$^{ 22}$Department of Physics and Astronomy, Tel Aviv University,
Tel Aviv 69978, Israel
\newline
$^{ 23}$International Centre for Elementary Particle Physics and
Department of Physics, University of Tokyo, Tokyo 113-0033, and
Kobe University, Kobe 657-8501, Japan
\newline
% ...
 
\section{Introduction}\label{sec:intro}
 
The measurement of the forward-backward asymmetries of heavy quarks,
\Aqq\ (q=b,c), in \eeqq\ events
provides an important test of the Standard Model. 
The \bb\ forward-backward asymmetry
provides one of the most precise determinations of 
the effective electroweak mixing angle 
for leptons \swsqeffl\ (assuming lepton universality).
This is of particular interest at the present time in view
of the nearly three standard deviation
difference between the average values of \swsqeffl\ derived
from quark forward-backward asymmetries at LEP 
on the one hand and 
from lepton forward-backward asymmetries
at LEP and the left-right asymmetry at SLD
on the other~\cite{ref:s02ew}.
 
The forward-backward asymmetry arises from the $\costh$ term
in the differential cross-section for \eeqq,
\begin{equation}
  \frac{\mathrm{d} \sigma} {\mathrm{d} \cos \theta} \propto 1 + \cos^2 \theta +
            \frac 8 3 \Aqq\ \cos \theta \, ,
\end{equation}
where $\theta$ denotes the angle between the outgoing quark and the
incoming electron flight directions, and where initial and final state 
radiation, quark mass effects and higher order terms have been neglected.
The asymmetry \Aqq\ is related to the vector, $\gv$, and
axial-vector, $\ga$, couplings of the \Zzero\ to the electron e and quark q.
At the peak of the \Zzero\ resonance and for the $s$-channel
\Zzero\ exchange process only, the pole asymmetry is given by
\begin{equation}
 \Afbzq \equiv \frac{3}{4} \cAe \cAq =
         \frac{3}{4}\ \frac{2 \gve/\gae }{ 1 + (\gve/\gae) ^2}\ 
                     \frac{2 \gvq/\gaq }{ 1 + (\gvq/\gaq) ^2} \, .
\end{equation}
In the Standard Model the couplings for any fermion f
are related to the fermion charge 
$Q_{\mathrm f}$ 
and its effective electroweak mixing angle $\sin^2\thwefff$ as follows:
\begin{equation}
\frac {\gvf}{ \gaf} = ( 1 -  {4 |Q_{\mathrm f}|}  \sin^2\thwefff ) \, .
\end{equation}
The values of $\sin^2\thwefff$ are all close to 0.25, so the value
of the asymmetry parameter for electrons, $\cAe$, 
is small, and varies rapidly with $\swsqeffl$, 
but the value of $\cAb$ is large, approximately 0.94, and
varies only slowly with $\sin^2\thweffb$. This 
results in a relatively large forward-backward asymmetry for 
\bb\ events, which is then 
very sensitive to \swsqeffl\ via $\cAe$.
  
This analysis uses hadronic \Zzero\ decays observed
by the OPAL detector at LEP to measure \Afbb\ and \Afbc. 
The event thrust axis is
used to estimate the primary quark direction.
Leptons produced in semileptonic
decays of b and c hadrons, usually referred to as prompt leptons, 
are used to
identify heavy flavour events, and their charge is used to 
distinguish between decaying quarks and antiquarks.
The asymmetry for \bb\ events is diluted by the effect of 
neutral B meson mixing~\cite{PDG}. 
This is quantified by
the average mixing parameter, \mix, which is the probability that a produced
b hadron decays as its antiparticle. The effect of B mixing is 
to reduce the observed asymmetry of \bb\ events
by a factor $(1 - 2 \mix)$ with respect to the asymmetry
without mixing.
The parameter \mix\ can be measured from the fraction of like-sign
lepton pairs in events with an identified lepton in each hemisphere
of the event.
The b and c asymmetries, and the mixing parameter
\mix, are therefore measured in a simultaneous fit to events with 
one or two identified leptons. The contributions of \bb\ and \cc\
events to the lepton samples are separated from each other and from background
by using several kinematic variables describing 
the lepton and its associated jet,
combined using neural network flavour separation algorithms.

This paper supersedes our
previous publication~\cite{ref:olasy}. Compared to the previous
paper, this
analysis benefits from the inclusion of data recorded after 1994, 
and from a reprocessing of the full OPAL data set with final
tracking algorithms and detector calibrations, which have
improved in particular the performance of the electron identification 
and flavour separation algorithms. 
The systematic uncertainties have been reduced due to better knowledge of
heavy flavour production and decay,
and several details of the fit have also been improved.
Similar analyses have been published by the other LEP
experiments~\cite{ref:alasy,ref:dlasy,ref:llasy}. 
Analyses identifying b quark events 
via the resolvable decay length of b 
hadrons~\cite{ref:ajet,ref:djasy,ref:ljet,ref:ojet}
also give precise measurements of the b quark 
forward-backward asymmetry. In addition,
the charm asymmetry has been measured 
using reconstructed charm 
hadrons~\cite{ref:adsac,ref:ddasy,ref:odsac}.

\section{Data sample and event simulation}\label{sec:data}

The OPAL detector is described in detail elsewhere 
\cite{opaldet,opalsi3d,opalsilep2}. 
The central tracking system is located inside a solenoid 
which provides a uniform axial magnetic field of 0.435~T along the
beam axis. The
beam axis coincides with the $z$-axis of the detector, with
the polar angle $\theta$ measured with respect to the direction
of the electron beam. The azimuthal angle $\phi$
is measured around the $z$-axis, and the radius $r$ is
the distance from the $z$-axis.
The innermost part of the tracking system
is a two-layer silicon micro-vertex detector. The
silicon detector is surrounded by
a vertex drift chamber, a large volume jet chamber with
159 layers of axial anode wires and a set of $z$ chambers
measuring the track coordinates along the beam direction. 
The jet chamber also provides a measurement of the rate of
energy deposition along the track, \dEdx.

Outside the magnet coil, the tracking system is surrounded by a
lead-glass electromagnetic calorimeter. The barrel region covers
the polar angle range
$|\cos\theta|<0.82$, and full acceptance to 
$|\cos\theta| < 0.98$ is provided by the endcaps.
The iron return yoke and pole pieces of the magnet are 
instrumented with streamer 
tubes and thin multiwire chambers to act as a hadronic calorimeter.
The calorimeters are surrounded by muon chambers.

Some flavour separation aspects of this analysis exploit the
long lifetimes of b and c hadrons, and therefore rely in
particular on the silicon microvertex detector. This detector was
first operational in 1991, providing measurements in the
$r$-$\phi$ plane only. In 1993 it was upgraded to measure tracks in both
$r$-$\phi$ and $r$-$z$ planes~\cite{opalsi3d}, and in 1996 the 
$\cos\theta$ coverage for at least one silicon measurement
was extended from $|\cos\theta|<0.83$ to $|\cos\theta|<0.93$~\cite{opalsilep2}.
The changing detector
performance with time was taken into account in the analysis. The resolution
of the vertex drift chamber is sufficient to ensure some flavour separation
even in data taken without silicon microvertex detector information.

Hadronic \Zzero\ decays were selected using standard criteria, 
as in~\cite{opalrb}.  All the LEP1 data collected at centre-of-mass
energies close to the peak of the 
\Zzero\ resonance between 1990 and 1995 were included,
together with \Zzero--peak data recorded for detector calibration
purposes during the higher energy LEP2 running between 1996 and 2000.
The thrust axis direction was calculated
using charged particle tracks and electromagnetic calorimeter clusters 
not associated
to any track. The polar angle of the thrust axis $\theta_{\mathrm{T}}$  
was required
to satisfy $|\cthr|<0.95$. 
Over four million hadronic events were selected. The exact
numbers are listed in Table~\ref{tab:events}.
As indicated in the table, some events 
were recorded at centre-of-mass energies above and below the \Zzero\ peak.
The bulk of these off-peak data 
were recorded in 1993 and 1995 at 
centre-of-mass energies approximately 1.8~GeV away from the peak.
The other off-peak centre-of-mass energy samples
from 1990 and 1991 are combined with 
these main samples, yielding measurements of 
the forward-backward asymmetries at three separate energy points.

\begin{table}[hbt]
\begin{center}
\begin{tabular}{l@{~$\langle \sqrt s \rangle = $~}r@{.}l@{~GeV~}|r@{~}c@{~}l|r@{~}l|r@{~}l|r@{~}l}
\mco{3}{c|}{Mean Energy}        & \mco{3}{|c|}{Selected}    & 
   \mco{4}{|c|}{Single lepton events} & \mco{2}{|c}{Dilepton} \\
\mco{3}{c|}{of \Zzero\ events} & \mco{3}{|c|}{\Zzero\ decays } & 
   \mco{2}{|c|}{Electrons} & \mco{2}{|c|}{Muons} & \mco{2}{|c}{events} \\ 
\hline
peak--2  & 89&51 &  &194&211 &  11&567 &  19&809 &  &321 \\
peak     & 91&25 & 4&079&047 & 239&505 & 410&877 & \phantom{00}6&014 \\
peak+2   & 92&95 &  &278&257 &  16&977 &  30&066 &  &394 \\
\end{tabular}
\end{center}
\caption{The number of events selected below, on and above the
\Zzero\ peak. The average centre-of-mass energies of the selected
\Zzero\ events, the numbers of \Zzero\ events, and the 
numbers of events with identified leptons are shown. Note that
dilepton events are only selected in a region of high b-purity.}
\label{tab:events}
\end{table}

Monte Carlo simulated events were generated using JETSET 7.4~\cite{jetset}
with parameters tuned by OPAL~\cite{jetsetopt}. The fragmentation function
of Peterson~\etal~\cite{fpeter} was used to describe the fragmentation of b
and c quarks. 
The semileptonic decay model of Altarelli \etal~\cite{ref:accmm}
with parameters fixed by CLEO, DELCO and MARKIII
data~\cite{ref:cleo,ref:delco,ref:markIII} was used to predict the
lepton momentum in the rest frame of  b and c hadrons.
The generated events were passed through a program
that simulated the response of the OPAL detector~\cite{gopal} and through
the same reconstruction algorithms as the data. 

\section{Lepton identification and flavour separation}

Leptons were identified in hadronic events using well established
algorithms~\cite{opalrb}. 
All tracks with momentum greater than 2~GeV 
and $|\cos \theta|<0.96$ were considered
as lepton candidates.

Electron candidate tracks were required to have at least 20
measurements of \dEdx\ from jet chamber hits. 
Electrons were then identified
using a neural network algorithm. The identification relies on 
ionisation energy loss (\dEdx) measured in the tracking chamber, 
together with spatial and
energy-momentum ($E/p$) matching between tracking and calorimetry. 
The neural network output was in the range 0 to 1, and
was required to be greater than 0.9 for electron candidates. Photon
conversions were rejected using another neural network 
algorithm~\cite{opalrb}. The efficiency of this selection for
genuine electrons, not considering electrons from photon conversions,
is 66\,\%. The purity is 73\,\%, where photon conversions
are included in the background.

Muons were identified by
requiring a spatial match between a track reconstructed in the tracking
detectors and a track segment reconstructed in the external muon chambers. 
Further rejection of kaons was achieved with 
loose \dEdx\ cuts~\cite{opalrb}. These requirements selected muons
with an efficiency of 74\,\% and a purity of 53\,\%. 

In this analysis, only 
leptons from the decay of a b or c hadron in a primary \bb\ or 
\cc\ event are considered as signal.
Any other genuine
electron or muon, and any hadron misidentified as a lepton,
is considered as background. With this definition,
about 11\,\% of the electron background
is from genuine electrons in 
light quark events, 49\,\% from misidentified hadrons
in any primary quark flavour event, and 40\,\% from
photon conversions. About 68\,\% of the 
muon background originates from pions, and 28\,\% from
kaons.

The relationship
between the lepton charge and the 
primary quark or antiquark from whose decay it
originated is vital for the asymmetry measurement. 
Leptons coming
directly from the weak decay of b hadrons are denoted \bl.
A negatively-charged
lepton comes from the decay of a hadron containing a b quark, and a positive
lepton from a b antiquark\footnote{Charge
conjugate decays are implied throughout this paper.}.
Electrons and muons from leptonic $\tau$ decays
where the $\tau$ lepton comes from a direct b decay, \btaul, have the same
sign correlation as the \bl\ events. Both lepton charges are possible
if a b hadron decays to a c hadron, which then decays
semileptonically. These decays are written as \bcl\ and \bcbarl,
according to the charm content of the intermediate meson, and are both
called cascade decays. In this analysis, 
any identified leptons from $\cc$ mesons,
e.g.\ $\mathrm{J/\psi}\rightarrow \ell^+\ell^-$ 
decays, are included with the cascades of the
appropriate sign. 
A neutral B meson may have mixed
before decay, so that a primary b quark decays as a b antiquark,
or vice versa. Leptons from these mixed mesons
are classified according to the decaying
b quark, and contribute to the asymmetry with the wrong sign
for their category, causing a reduction in the observed
\bb\ asymmetry by a factor of $(1 - 2\mix)$.
Leptons from the decay of c hadrons produced in
\cc\ events, $\cl$, have the opposite sign correlation to direct $\bl$
decays; a primary c quark gives a positive lepton. 

As in the previous analysis~\cite{ref:olasy},
two neural networks \cite{jetnet} denoted \NETb\ and \NETc\ were used to
separate \bl\ and \cl\ decays from each other, from
the cascade decays, \bcl\ and \bcbarl\ which dilute the observed
forward-backward asymmetries, and from backgrounds. The
networks were retrained to take full advantage of the 
improved detector calibrations. Several of the network input
variables refer to the jet containing the lepton track.
The same tracks and clusters used to define the event thrust
axis were combined into jets using a cone algorithm~\cite{jetcone} with a
cone half-angle of 0.55~rad  and a minimum jet energy of 5.0~GeV. 
The transverse momentum, \pt, of each track was defined relative
to the axis of the jet containing it, where the jet axis was calculated
excluding the momentum of the track. 
A lepton sub-jet was defined as in~\cite{ref:olasy}.
The sub-jet includes particles that
are nearer to the lepton than to the jet axis, and is a 
measure of the lepton isolation. 

\begin{figure}
\setlength{\epsfxsize}{0.60\hsize}
\epsfbox{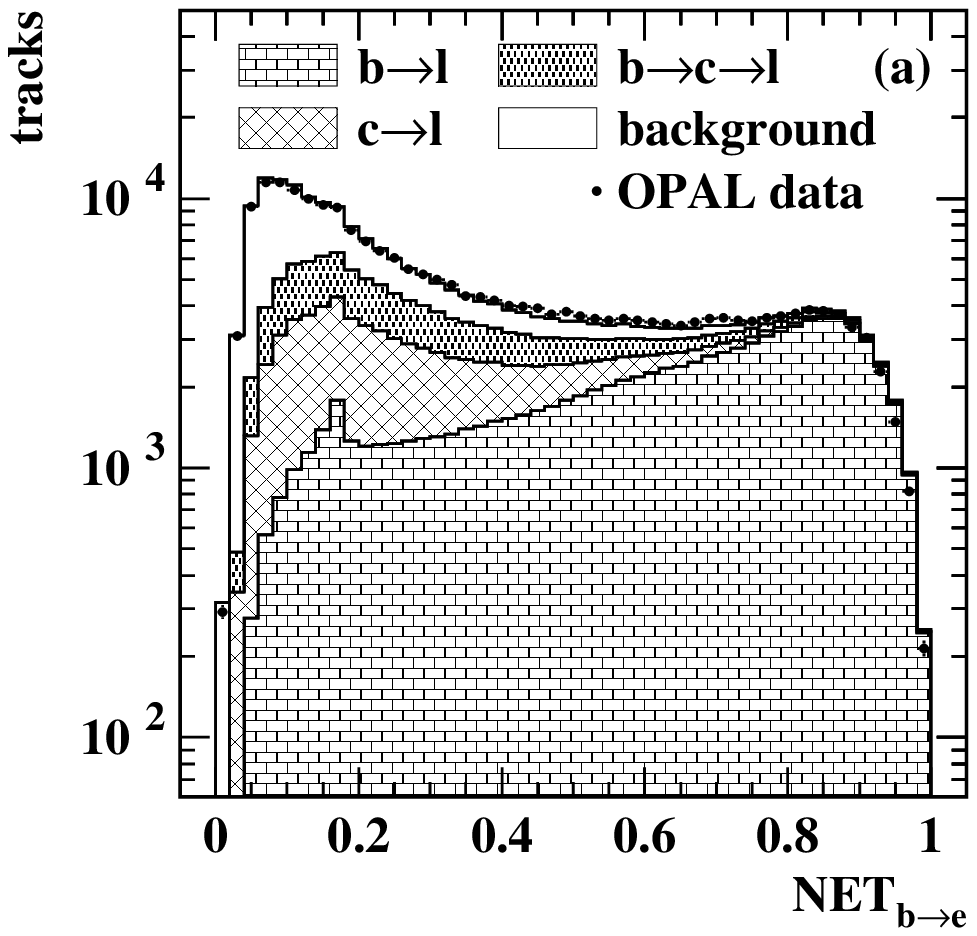}
\hspace{-18mm}
\setlength{\epsfxsize}{0.60\hsize}
\epsfbox{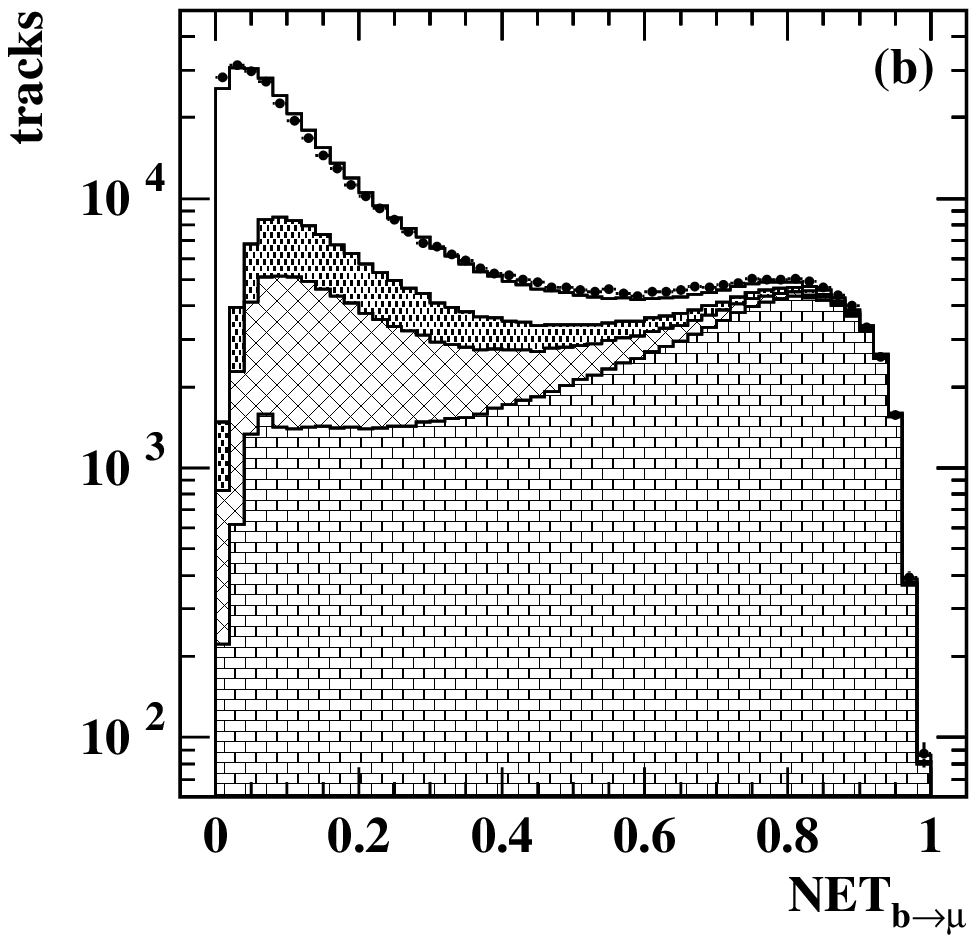}
\vspace{-10mm}

\setlength{\epsfxsize}{0.60\hsize}
\epsfbox{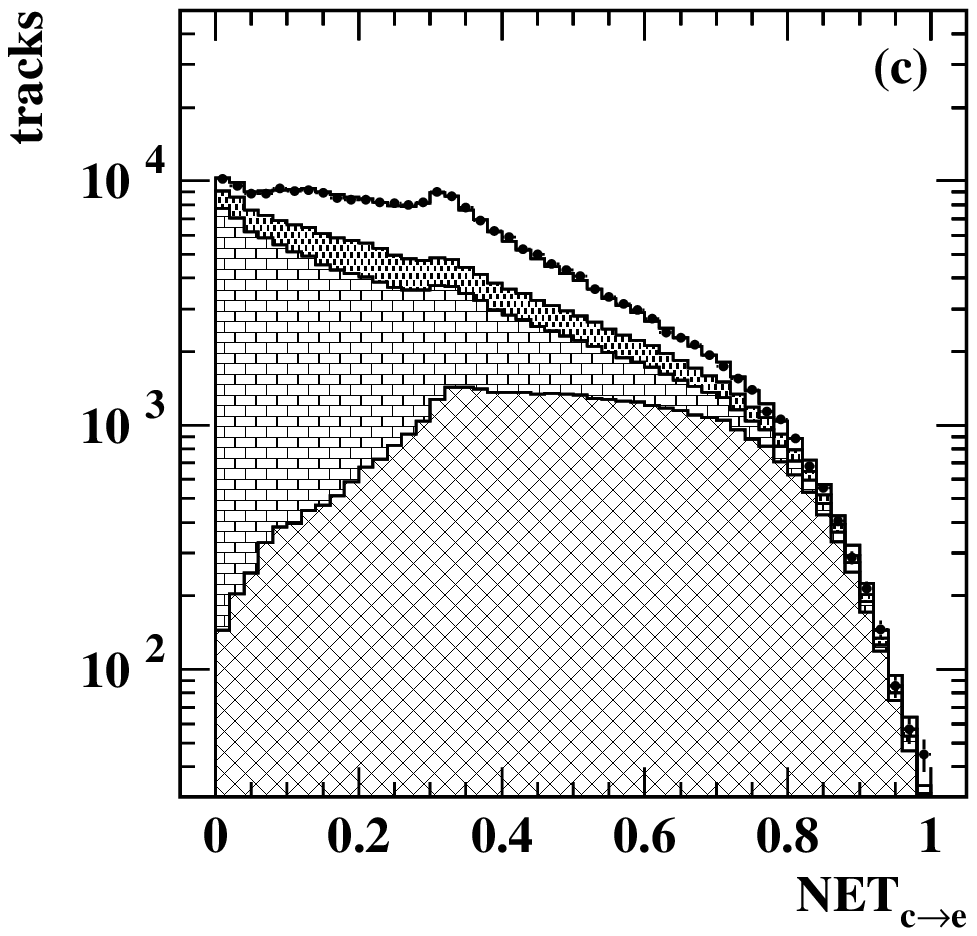}
\hspace{-18mm}
\setlength{\epsfxsize}{0.60\hsize}
\epsfbox{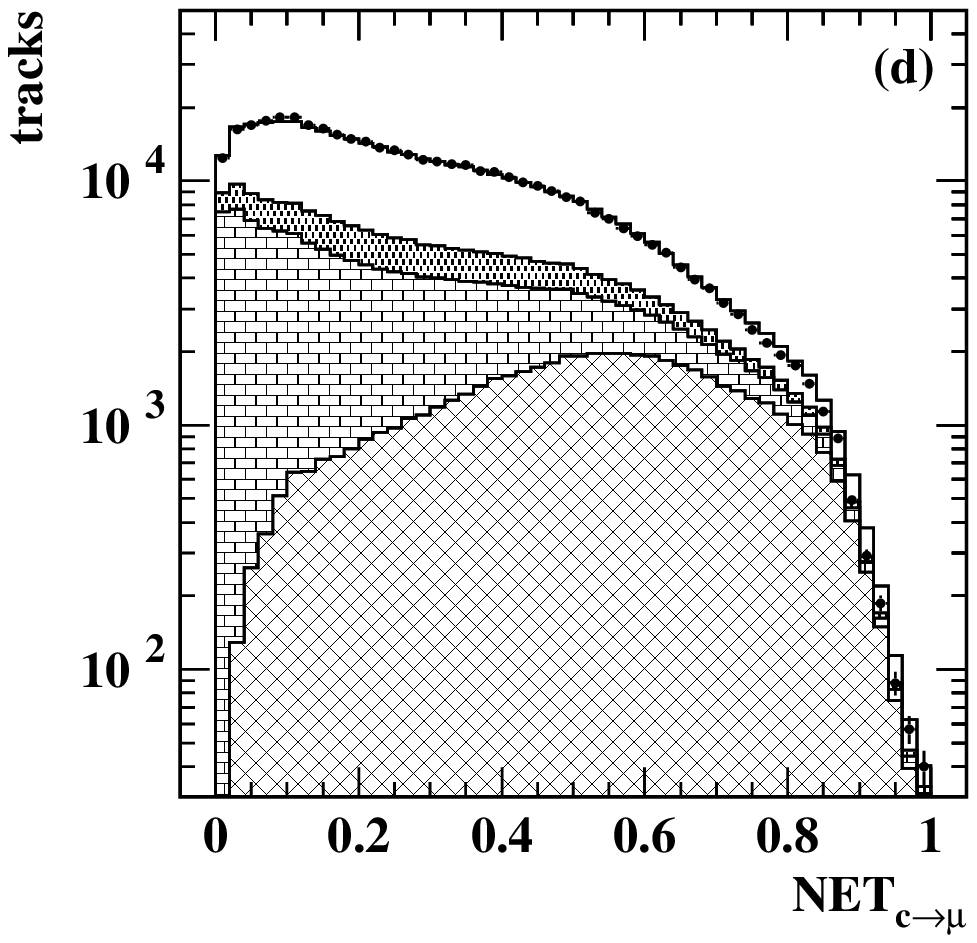}

\caption{\label{fig:netdist}Outputs of the neural networks designed to
select (a) $\rm b\rightarrow e$, (b) $\rm b\rightarrow\mu$, (c) 
$\rm c\rightarrow e$ and (d) $\rm c\rightarrow\mu$ decays. The data
are shown by the points with error bars, and the expected contributions
from different lepton sources by the hatched histograms.}
\end{figure}

The first network, \NETb, was trained to distinguish between
\bl\ events and all other categories. The input variables
were the momentum $p$ and transverse momentum $p_t$ of the
lepton candidate, 
the energy of the lepton sub-jet, \esub,
 the total visible energy of the jet (calculated using all
tracks and unassociated calorimeter clusters comprising the jet), \jte, and
the scalar
sum of the transverse momentum of all tracks within the
 jet, \jspt.
Separate networks
were trained for electrons and for muons. For the electron
nets, two extra variables were included, namely
the outputs of the electron
and conversion neural networks.
The distributions of the NETb output in the OPAL data are shown 
in Figure~\ref{fig:netdist}(a) and (b), together with the predictions from
Monte Carlo; at high NETb values, the separation of $\bl$ 
events from all other lepton sources is clearly visible. For example, requiring
NETb\ $>0.7$ gives a sample 89\,\% pure in \bl\ decays, with 
3\,\% \bcl\ and 4\,\% \cl, 
whilst retaining
45\,\% of all \bl\ decays with an identified lepton (averaged over 
electrons and muons). The Monte Carlo gives a reasonable description of the
data, and the effect of the small discrepancies visible is discussed in 
Section~\ref{sec:other}.
 
The second network, \NETc, was trained to distinguish
\cl\ events from all other categories, including \bl. The
network used all the \NETb\ variables, including the 
electron and conversion neutral network outputs for 
electron candidates, together with the following
three quantities:
the decay length significances, \siglb, of the jet containing the
lepton (jet 1) and of the most energetic of the other jets in the event
(jet 2), and
the impact parameter significance of the lepton with
respect to the primary vertex, \sigb. 
The decay length significance is the distance between
the primary vertex and a secondary vertex, reconstructed from a subset of
the tracks in the jet
using the algorithms described in~\cite{opalrb},
divided by the error on the decay length. The impact parameter
significance is the distance of closest approach of a track
to the primary vertex, divided by its error.
The distributions of the NETc output in the OPAL data are shown 
in Figure~\ref{fig:netdist}(c) and (d), together with the predictions from
Monte Carlo; again the separation of \cl\ events from 
other leptons is clearly visible. Requiring NETc\ $>0.7$ gives a sample
59\,\% pure in \cl\ decays, with 12\,\% \bl and 6\,\% \bcl,
whilst retaining 19\,\% of all \cl\ decays with
an identified lepton.

The values of NETb and NETc were evaluated for all lepton candidates.
When more than one lepton was identified in an event, the candidates
were ranked according to the value of \NETb.
If the best two lepton candidates in an event satisfied $\NETb>0.6$,
and were in opposite thrust hemispheres, then 
both candidates were retained, and this was classified as a dilepton event.
Otherwise only the best lepton candidate was considered, and the 
event was classified as a single lepton event. The numbers
of single and dilepton events selected
are given in Table~\ref{tab:events}.

\section{Fit method and results}\label{s:fitres}

The values of the asymmetries and \mix\ were determined using a simultaneous
fit to the observed numbers of single lepton events as a function of 
$\cthr$, NETb and NETc, and the numbers of dilepton events as a function
of $\cthr$.
The thrust axis of each event was used to estimate the direction
of the primary quark, and the charge $Q_{\ell}$ of the identified leptons
was used to distinguish between the quark and antiquark direction.

The total likelihood to be maximized is the product:
\begin{equation}
  \cal{L}  =   \cal{L}_{\rm single\,\, e} \times 
               \cal{L}_{\rm single\,\, \mu} \times
               \cal{L}_{\rm double} \,\,. \label{eqn:totl}
\end{equation}

The single and double lepton likelihood terms are discussed in more detail
below.

\subsection{Single lepton likelihood}

For single lepton events, 
the observable $y = -Q_{\ell} \cthr $ was used to classify events as 
forward, with $y>0$, or backward, with $y<0$.
The observed forward-backward asymmetry was then examined in
bins of \NETb, \NETc\ and $|y|$. The fit considers 
four classes of leptons according to the lepton charge and
primary quark flavour:
\begin{enumerate}
\item \bl, \btaul\ and \bcbarl
\item \bcl
\item \cl
\item background
\end{enumerate}
Direct \bl\ decays and the other decays in class (1) contribute
to the observed asymmetry with the opposite sign to classes
(2) and (3). The asymmetry of 
events coming from b decays is scaled by a 
factor $(1-2\mix)$ for category (1) decays, and 
$(1- 2 \overline{\eta}\,  \mix)$ for category (2), to account
for the effects of neutral B meson mixing.
The additional
correction factor $\oeta$ takes into account that 
the two samples include different 
fractions of different species of b hadron: $\Bzero$, $\Bplus$,
$\Bzeros$, b-baryons etc.,
due mainly to the different semileptonic
branching ratios of $\Dzero$ and $\Dplus$
mesons.
The value of $\oeta$ is set to 1.083, 
evaluated from the Monte Carlo simulation.
No equivalent correction factor was used for the small fraction
of $\bcbarl$ events included in category (1).
The fractions of each single lepton category in the full data sample
are given in Table~\ref{tab:fracsingle}. Note that these 
are fitted fractions. The fit adjusts the overall rate 
of background in the sample.

\begin{table}[hbt]
\begin{center}

\begin{tabular}{ll|r@{.}l|r@{.}l}
& & \mco{2}{|c|}{electrons} & \mco{2}{|c}{muons} \\ \hline
(1) & \bl            & 35&6 \% & 24&6 \% \\
    & \bcbarl        &  2&8 \% &  2&3 \% \\
    & \btaul         &  1&1 \% &  0&7 \% \\
(2) & \bcl           & 12&6 \% &  9&7 \% \\
(3) & \cl            & 19&3 \% & 14&9 \% \\
(4) & background     & 28&6 \% & 47&8 \% \\ 
\end{tabular}
\end{center}

\caption{The composition of the single lepton
samples. The first three lines are combined into
category (1) in the fit. The fractions given correspond
to the fractions $f'_i$ after the fit.}
\label{tab:fracsingle}
\begin{center}
\begin{tabular}{l|r@{.}l@{\,\% ~}|r@{.}l@{\,\% ~}|r@{.}l@{\,\% ~}}
& \mco{2}{|c|}{ee} & \mco{2}{|c|}{$\mu \mu$} & \mco{2}{|c}{e$\mu$} \\ \hline
(1)(1) & 86&9 & 86&3 & 86&4 \\
(1)(2) & 10&3 &  8&5 &  9&6 \\
(2)(2) &  0&3 &  0&2 &  0&3 \\
(3)(3) &  0&4 &  0&4 &  0&5 \\
(1)(4) &  1&8 &  4&0 &  2&8 \\
others &  0&3 &  0&6 &  0&4 \\ \hline
total number & \mco{2}{|r|}{1~308} & \mco{2}{|r|}{2~067} & \mco{2}{|r}{3~354} \\
\end{tabular}
\end{center}
\caption{Fractions of events in the most important categories in the 
double lepton sample, together with the total number
of ee, $\mu\mu$ and e$\mu$ events.}
\label{tab:fracdouble}
\end{table}

\begin{table}[htb]
\begin{center}
\begin{tabular}{l|r@{.}l@{$\pm$}r@{.}l
                |r@{.}l@{$\pm$}r@{.}l
                |r@{.}l@{$\pm$}r@{.}l}
& \mco{12}{|c}{$A_{\mathrm{FB}}$ (\,\%) (ZFITTER  prediction) } \\
$\langle \sqrt{s} \rangle$ (GeV) & 
   \mco{4}{|c|}{89.51} &
   \mco{4}{|c|}{91.25} &
   \mco{4}{|c}{92.95} \\ \hline
{\mbox{$\rm s\overline{s}$}} & 5&95 &  5&0 & 9&64 & 1&3 &11&89 & 5&0 \\
{\mbox{$\rm u\overline{u}$}} &-3&08 & 20&0 & 6&32 & 7&3 &12&07 & 20&0 \\
{\mbox{$\rm d\overline{d}$}} & 5&96 & 10&0 & 9&64 & 3&7 &11&89 & 10&0 \\ 
\end{tabular}
\caption{The values of the forward-backward 
asymmetries for light flavours taken from ZFITTER, and the 
variations used for calculating the background asymmetry.}
\label{tab:bkgasyinput}
\vspace*{0.5cm}
\begin{tabular}{l|r@{.}l|r@{.}l|r@{.}l|r@{.}l}
& \mco{4}{|c|}{electrons}
& \mco{4}{|c}{muons} \\ 
& \mco{2}{|c|}{$f^{\rm q\overline{q}}_{\rm back}$}
& \mco{2}{|c|}{$c^{\rm q\overline{q}}_{\rm dilute}$}
& \mco{2}{|c|}{$f^{\rm q\overline{q}}_{\rm back}$}
& \mco{2}{|c}{$c^{\rm q\overline{q}}_{\rm dilute}$} \\ \hline
{\mbox{$\rm b\overline{b}$}} & 0&139 & 0&028  & 0&192 & 0&093 \\
{\mbox{$\rm c\overline{c}$}} & 0&146 & -0&104 & 0&178 & -0&043 \\
{\mbox{$\rm s\overline{s}$}} & 0&214 & 0&025  & 0&243 & 0&103 \\
{\mbox{$\rm u\overline{u}$}} & 0&220 & -0&114 & 0&169 & -0&123 \\
{\mbox{$\rm d\overline{d}$}} & 0&218 & 0&107  & 0&218 & 0&083 \\
\end{tabular}
\end{center}
\caption{The fractions of background coming from each primary quark
flavour for electrons and muons, 
and the dilution factors by which the primary quark
forward-backward asymmetry are scaled to evaluate the background
asymmetry.}
\label{tab:bkgasyfrac}
\end{table}

The likelihood for single-lepton events has two terms:
\begin{eqnarray}
\nonumber
{\cal L}_{\rm   single\,\,\ell }  &=&  \prod_{\NETb,\NETc} \; \prod_{ |y| }
  \frac{(\nf+\nb)!}{\nf! \; \nb!}
      \;  \left( \frac{1+\asy}{2} \right)^{\nf}
      \;  \left( \frac{1-\asy}{2} \right)^{\nb} \\
&\times  &    \prod_{\NETb,\NETc} \phantom{\prod_{ |y| }}
 \frac{1}{\sqrt{2 \pi} \sigma_{m}} \;
      \exp \left(-\frac{(\nt-m)^2}{2 \sigma_{m}^2}\right).
\label{equ:lik}
\end{eqnarray}
The first term is a product over bins
of NETb, NETc and $|y|$ of 
binomial probabilities for the number of forward and backward
events observed in the data in each bin, \nf\ and \nb, depending on the
asymmetry \asy\ expected in the bin.
The fit used 10 equally spaced
bins in each of NETb, NETc and $|y|$.
The second term 
constrains the total number
of events \nt\
in a given NETb-NETc bin (with no binning in $|y|$)
to the expected number, $m$,
and allowed the overall normalisations  of  the
electron and muon backgrounds to be determined from the data,
reducing the overall uncertainty.

The probability for an event to be forward is $(1+\asy)/2$
where \asy\ is the expected forward-backward  asymmetry in the
bin considered:
\begin{equation}
\asy (\NETb,\NETc, |y|) =  
\sum_{i=1}^{4} \fraci' \rho_i(\NETb,\NETc,|y|)
\asy^i\ \frac{8}{3} \frac{|y|}{1+y^2} \, .
\label{eqn:asy}
\end{equation}
In this expression, $\fraci'$ denotes
the fraction of leptons in class $i$, and
$\rho_i$ the normalised distribution of leptons
from this class in bins of NETb, NETc and $|y|$,
which is taken from Monte Carlo simulation.
The source fractions $\fraci'$ 
are derived from the Monte Carlo fractions $\fraci$.
However, the 
fraction of background $f_4'$ is a free parameter
in the fit, and the fractions of prompt sources with $i=1,2,3$ are
given by $\fraci' = (1 - f_4') \fraci/(f_1+f_2+f_3)$.
In this way, 
the relative rates of the prompt leptons are
fixed by the Monte Carlo simulation, and
the fractions satisfy $\sum_i \fraci' = 1$.
The nominal asymmetries $\asy^i$ for each class are:
\begin{equation}
\begin{array}{lccl}
 \asy^1 & = &   & ( 1-2 \mix) \Afbb  \\
 \asy^2 & = & - & ( 1- 2 \oeta\,\mix) \Afbb  \\
 \asy^3 & = & - & \Afbc \\
 \asy^4 & = &   & \asy^{\mathrm{Background}} \\
\end{array}
\end{equation}
The background asymmetry $\asy^{\mathrm{Background}}$ also depends on 
the primary quark asymmetries. It is expressed as a sum
over quark flavours:
\begin{equation}\label{e:bgeq}
\asy^{\mathrm{Background}} = \sum_{\mathrm{q}}
f^{\qq}_{\mathrm{back}} c^{\qq}_{\mathrm{dilute}} \asy^{\qq} \, .
\end{equation}
This introduces a very weak additional dependence on the 
fitted values of the b and c asymmetries. The central values of the 
light quark asymmetries are taken from the predictions of 
ZFITTER 6.36~\cite{zfitter},
and are listed in Table~\ref{tab:bkgasyinput}. 
The quoted uncertainties
in these asymmetries are taken from measurements of
the strange quark asymmetry by DELPHI~\cite{delphistrange} and of
the light quark asymmetries by OPAL~\cite{opallight}. These
measurements are consistent with the ZFITTER expectations,
with relatively large statistical errors.
The fractions $f^{\qq}_{\mathrm{back}}$ of each quark flavour 
contributing to the background are taken from the Monte Carlo prediction,
as are the dilution factors $c^{\qq}_{\mathrm{dilute}}$ which 
take into account the fraction of the primary quark
asymmetry that is seen in background events of this flavour. 
The fractions and dilutions are listed in Table~\ref{tab:bkgasyfrac}.
Some contributions to the background, for example for photon conversions,
have zero forward-backward asymmetry. Others, in particular kaons
in \antibar{s}\ events, inherit a significant fraction of
the primary quark asymmetry. The background asymmetry actually
varies as a function of \NETb\ and \NETc, but this effect is neglected in
the fit as discussed in Section~\ref{s:systbg} below.

In the second term in the single lepton likelihood, 
the number of events, $m$, expected in a bin
of NETb and NETc is calculated by:
\begin{equation}
m (\NETb,\NETc ) = N^\ell_{\rm Data} \sum_{i=1}^{4} \fraci' 
\overline{\rho_i}(\NETb,\NETc) \, ,
\end{equation}
where $N^{\ell}_{\rm Data}$ denotes the total number of
single lepton events in the data ($\ell=e,~\mu$),
and $\overline{\rho_i}$ is the normalised distribution 
in bins of NETb and NETc for leptons in class $i$.
The observed number of events in a bin, \nt, is
assumed to be Gaussian distributed with mean $m$ and
standard deviation
$\sigma_{m}$ computed as
$ \sigma_{m}^2 = m + (\delta_{m}^{\rm
MC})^2$,  where $\delta_{m}^{\rm MC}$ is the uncertainty
on the expected number $m$ of events in this bin due to the limited
Monte Carlo statistics.

\subsection{Double lepton likelihood}

The double lepton likelihood is a product in bins of $|y|$
of multinomial
probabilities for the event to be forward, $\pf$,
backward, $\pb$, or same-sign, $\ps$. Both leptons have
$\NETb>0.6$, and no further subdivision
in NETb or NETc is made. The composition of the
double lepton sample is given in Table~\ref{tab:fracdouble}.
The ee, e$\mu$ and $\mu\mu$ events
are all considered together:
\begin{equation}
{\cal L}_{\rm double} =  \prod_{ |y| }
  \frac{(\nf+\nb+\ns)!}{\nf! \; \nb! \; \ns!}
      \;  \left( \pf \right)^{\nf}
      \;  \left( \pb \right)^{\nb}
      \;  \left( \ps \right)^{\ns}   \, .
\end{equation}
In this case, \nf, \nb\ and \ns\ are the number of 
forward, backward and same-sign events observed in the data
in this $|y|$ bin.
A dilepton event with
opposite charge leptons is called forward or backward
according to the direction of the negative lepton. The
probability of such an event being forward or backward usually
depends on the forward-backward asymmetry of the dilepton category, $ij$,
where the two indices refer to the categories $i=1-4$ of the two leptons.
The fraction of same-sign dilepton events depends only on
the mixing parameter $\mix$, and not on the asymmetries.
Note that the fraction of forward and backward events with one
lepton from category 1 and the other from category 2 is also
independent of the forward-backward asymmetry, and only
depends on the mixing parameter. This is because in the absence of
mixing these events would all be same sign, losing the 
information about the asymmetry. If they are of opposite sign,
without knowing which lepton comes from the mixed meson,
the asymmetry information cannot be recovered.

Only certain categories $ij$ are possible.
Any flavour event can give a background lepton, but lepton classes
1 and 2 can only come from \bb\ events, and class 3
only from \cc\ events. The possible dilepton classes are therefore:
$ij=11,12, 22, 33, 14, 24, 34, 44$.
The overall probabilities for 
forward, backward or same sign events are given by 
a sum over the possible classes $ij$:
\begin{equation}
p_X = \sum_{ij} \fracij(|y|) p_X^{ij}(|y|)
\end{equation}
Since the contribution of background
leptons to the double-lepton sample
is small (see Table~\ref{tab:fracdouble}), the
fractions \fracij\ are taken directly
from the Monte Carlo simulation, without being corrected for the rates
of background leptons fitted in the single lepton sample. Similarly, the
small residual asymmetries of the background leptons are neglected. 
Writing $Y = 8|y|/3(1+y^2)$, the 
probabilities $p_{\mathrm{F,B,S}}^{ij}$ 
for an event to be forward, backward or same-sign for each dilepton
category are given by:
\begin{eqnarray}
\pf^{11}&=&\hlf  (1-2\mix+2\mix^2 + (1-2\mix)  \Afbb  Y) \nonumber \\ \nonumber
\pb^{11}&=&\hlf  (1-2\mix+2\mix^2 - (1-2\mix)  \Afbb  Y) \\ \nonumber
\ps^{11}&=&2  \mix  (1-\mix)                             \\ \nonumber
\pf^{12}&=&\hlf\oeta\,\mix  (1-\mix) + \hlf\mix  (1-\oeta\,\mix) \\ \nonumber
\pb^{12}&=&\hlf\oeta\,\mix  (1-\mix) + \hlf\mix  (1-\oeta\,\mix) \\ \nonumber
\ps^{12}&=&(1-\mix)(1-\oeta\,\mix) + \oeta\,\mix^2           \\ \nonumber
\pf^{22}&=&\hlf  (1-2\oeta\,\mix + 2(\oeta\,\mix)^2 - (1-2\oeta\,\mix)  \Afbb  Y)\\ \nonumber
\pb^{22}&=&\hlf  (1-2\oeta\,\mix + 2(\oeta\,\mix)^2 + (1-2\oeta\,\mix)  \Afbb  Y)\\ \nonumber
\ps^{22}&=&2  \oeta\,\mix  (1-\oeta\,\mix) \\ \nonumber
\pf^{33}&=&\hlf  (1 - \Afbc  Y) \\ \nonumber
\pb^{33}&=&\hlf  (1 + \Afbc  Y) \\ \nonumber
\ps^{33}&=&0                   \\ 
\pf^{14}&=&\qrt  (1-\mix)  (1 + \Afbb  Y) + \qrt  \mix  (1 - \Afbb  Y)\\ \nonumber
\pb^{14}&=&\qrt  (1-\mix)  (1 - \Afbb  Y) + \qrt  \mix  (1 + \Afbb  Y)\\ \nonumber
\ps^{14}&=&\hlf                                              \\ \nonumber
\pf^{24}&=&\qrt  (1-\oeta\,\mix)  (1 - \Afbb  Y) + \qrt  \oeta\,\mix  (1 + \Afbb  Y) \\ \nonumber
\pb^{24}&=&\qrt  (1-\oeta\,\mix)  (1 + \Afbb  Y) + \qrt  \oeta\,\mix  (1 - \Afbb  Y) \\ \nonumber
\ps^{24}&=&\hlf \\ \nonumber
\pf^{34}&=&\qrt  (1 - \Afbc  Y) \\ \nonumber
\pb^{34}&=&\qrt  (1 + \Afbc  Y) \\ \nonumber
\ps^{34}&=&\hlf \\ \nonumber
\pf^{44}&=&\qrt \\ \nonumber
\pb^{44}&=&\qrt \\ \nonumber
\ps^{44}&=&\hlf \nonumber
\end{eqnarray}

\subsection{Results}\label{s:res}

The data are divided into three separate energy bins whose mean 
energies are shown in Table~\ref{tab:events}.
The asymmetries were fitted simultaneously at all energy points, the overall
likelihood being the product of the likelihood in Equation~\ref{eqn:totl}
for each point. Therefore, the fit has 
nine free parameters: the asymmetries \Afbb\ and  \Afbc\ at three energy 
points, together with values common to all three energy points for
the mixing parameter \mix, and the electron and muon background
fractions, $f'_4$.
Data from all years were fitted simultaneously, with the various fractions and
(NETb, NETc) distributions determined from an appropriate mix of Monte Carlo
events with different simulated detector configurations,
taking into account the changes in the performance of
the OPAL detector over time.

\begin{figure}
\begin{center}
\setlength{\epsfxsize}{0.98\hsize}
\mbox{\hspace*{1cm}\epsfbox{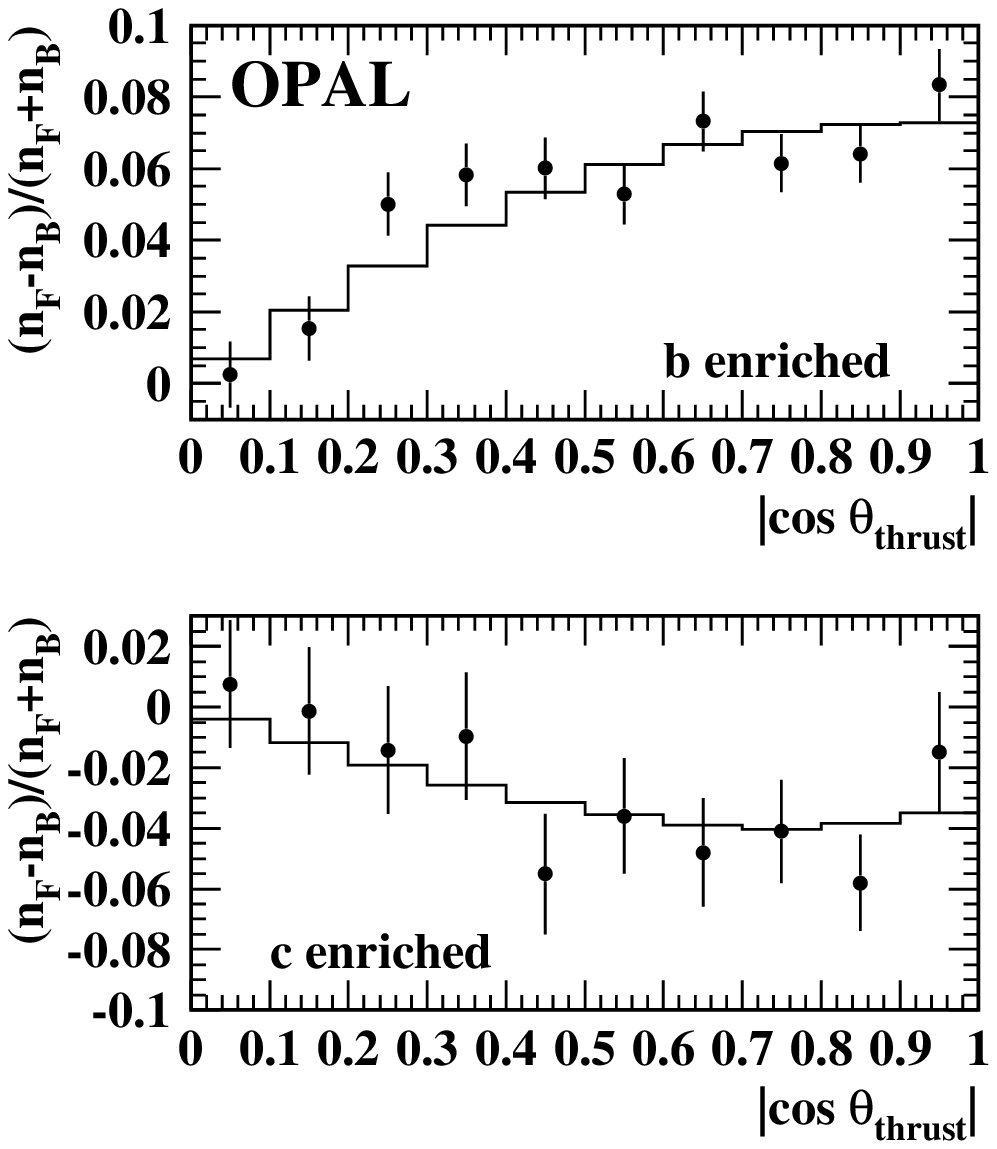}}
\end{center}

\caption{\label{fig:fit}
An illustration of the fit results for (a) a b enriched region and
(b) a c enriched region of NETb-NETc space. The asymmetry observed
in the data in each bin of $|y|$ is compared with the expectation
calculated from the full fit. The errors are purely statistical.
}
\end{figure}

The result of the fit is illustrated in Figure~\ref{fig:fit}. 
For this plot, two regions
of NETb-NETc were selected, one 93\,\% pure in \bl\ events and the other
59\,\% pure in \cl\ events. The asymmetry observed
in the data in bins of $|y|$ and the predicted asymmetry according
to Equation~\ref{eqn:asy} are shown. The predicted asymmetry is
calculated with the fitted results.
The sign of the observed asymmetry
is clearly different in the two regions.

A small correction is applied to the fitted asymmetries to correct for the
effects of gluon radiation from the primary quark pair and the approximation 
of the original quark direction by the experimentally measured thrust axis
\cite{qcdcor}. The effects of gluon radiation have been calculated to second 
order in $\alpha_s$ using the parton level thrust axis to define the asymmetry 
\cite{qcdcalc},
and the translation from the parton level to the hadron level thrust axis 
(defined using all final state particles without detector effects)
has been determined using Monte Carlo hadronisation models \cite{qcdcor}.
However, the detector acceptance and fitting method introduce additional
biases which reduce the sensitivity of the analysis to these QCD corrections.
Therefore, the final correction applied to the fitted asymmetries is determined
by comparing the asymmetries fitted on large samples of Monte Carlo 
simulated events with the true quark level asymmetries, after
scaling the hadron level thrust axis QCD corrections in the Monte Carlo to 
the theoretical values \cite{lephfew}.
This procedure also accounts for any other residual biases in the fit,
for example from the treatment of the background asymmetry.
Combining all effects, the raw fitted asymmetries were scaled 
by factors of
$1.0050\pm 0.0063\pm 0.0050$ for b quarks and $1.0117\pm 0.0063\pm 0.0062$ 
for c quarks
to determine the quark level results, where the first errors
result from theoretical uncertainties \cite{qcdcor,lephfew}
and the second from limited Monte Carlo statistics.

After correcting for QCD effects, the results are:
\begin{center}
\begin{tabular}{@{\Afbb = (}r@{.}l@{$\,\pm\,$}r@{.}l@{$\,\pm\,$}r@{.}l@{)\,\%\ \ \ 
                  \Afbc = (}r@{.}l@{$\,\pm\,$}r@{.}l@{$\,\pm\,$}r@{.}l@{)\,\%\ \ \ 
                  at $ \langle \protect\sqrt s \rangle = $ }r@{.}l@{}}
4&7 & 1&8 & 0&1 & $-$6&8 & 2&5 & 0&9 & 89&51 GeV,\\
9&72 & 0&42 & 0&15 & 5&68 & 0&54 & 0&39 & 91&25 GeV,\\
10&3 & 1&5 & 0&2 & 14&6 & 2&0 & 0&8 & 92&95 GeV.\\
\end{tabular}
\end{center}
For the average B mixing parameter, a value of:
$$
\mix = (13.12\pm 0.49 \pm 0.42)\,\%
$$
is obtained. In each case the first error
is statistical and the second systematic. The evaluation of 
the systematic errors is described in the following section.
The statistical correlations between the results are given
in Table~\ref{tab:correlations}.
The background levels in the electron and muon samples were 
fitted to be $(89.4\pm 0.6)$\,\% and $(94.4\pm 0.3)$\,\% of the rates
estimated from  Monte Carlo simulation, 
where the quoted errors are statistical only. These 
values are consistent with the
known uncertainties in modelling the lepton background levels
\cite{opalrb}.

\begin{table}[hbt]
\begin{center}
\begin{tabular}{l|r@{}c@{}l|r@{}c@{}l|r@{}c@{}l} 
{\bf peak} & 
\mco{3}{|c|}{\Afbb} &
\mco{3}{|c|}{\Afbc} &
\mco{3}{|c}{\mix} \\ \hline 
\Afbb & 1&.&00 & 0&.&17 & 0&.&30 \\ 
\Afbc & && & 1&.&00 & 0&.&01 \\
\mix  & && & && & 1&.&00 \\
\end{tabular} 
\hspace*{0.5cm}
\begin{tabular}{l|r@{}c@{}l|r@{}c@{}l|r@{}c@{}l} 
{\bf peak$-$2} & 
\mco{3}{|c|}{\Afbb} &
\mco{3}{|c|}{\Afbc} &
\mco{3}{|c}{\mix} \\ \hline 
\Afbb & 1&.&00 & 0&.&18 & 0&.&03 \\ 
\Afbc & && & 1&.&00 & 0&.&00 \\
\mix  & && & && & 1&.&00 \\
\end{tabular} 
\hspace*{0.5cm}
\begin{tabular}{l|r@{}c@{}l|r@{}c@{}l|r@{}c@{}l} 
{\bf peak$+$2} & 
\mco{3}{|c|}{\Afbb} &
\mco{3}{|c|}{\Afbc} &
\mco{3}{|c}{\mix} \\ \hline 
\Afbb & 1&.&00 & 0&.&17 & 0&.&10 \\ 
\Afbc & && & 1&.&00 & 0&.&00 \\
\mix  & && & && & 1&.&00 \\
\end{tabular} 
\end{center}
\caption{Statistical
correlation  matrices at each centre-of-mass energy using the entire
(1990--2000) data sample. The correlations with the 
fitted background fractions
are negligibly small ($<0.01$) and are not given here.}
\label{tab:correlations}
\end{table}

\section{Systematic uncertainties}

Systematic uncertainties result from a number of sources, including 
modelling of b and c hadron production and  decay, external branching 
ratio inputs, 
background uncertainties and the performance of the OPAL detector.
For the first two sources, the proposals of the
LEP Heavy Flavour Electroweak Working Group have been 
adopted~\cite{lephfew}. Where errors are only assessed by a 
variation in one direction, for example to an alternative model,
or when slightly 
asymmetric errors are assessed, the
larger deviation is taken as $1 \sigma$. The systematic
errors are summarised in Tables~\ref{tab:systpeak} and~\ref{tab:systoff} 
and discussed in more detail below. The signs of the errors indicate
the direction of change in the result when the corresponding quantity is
varied.

\begin{table}[p]
\begin{center}
\begin{tabular}{l|r@{.}l|r@{.}l|r@{.}l}
& \mco{2}{|c|}{\Afbb (\%)} 
& \mco{2}{|c|}{\Afbc (\%)} 
& \mco{2}{|c}{\mix (\%)} \\ \hline
Fitted Value & 9&716 & 5&683 & 13&121 \\
Statistical error & $\pm$0&418 & $\pm$0&542 & $\pm$0&485 \\
Systematic error & $\pm$0&150 & $\pm$0&386 & $\pm$0&421 \\ \hline  
\mco{7}{c}{ } \\
\mco{7}{c}{\bf Sources of systematic errors} \\ \hline
\bl\ semileptonic decay model            & $\pm$0&011 & $\mp$0&101 & $\pm$0&175 \\
\cl\ semileptonic decay model            & $\pm$0&064 & $\mp$0&036 & $\mp$0&286 \\
$\mxeb$ $\pm$0.008                       & $\mp$0&012 & $\mp$0&008 & $\mp$0&074 \\
$\mxec$ $\pm$0.008                       & $\pm$0&054 & $\mp$0&024 & $\pm$0&010 \\
{\bf Total models}                       &      0&085 &      0&110 &  0&344 \\
\hline
BR(\bl) (10.65 $\pm$0.23)\,\%            & $\mp$0&020 & $\pm$0&104 & $\pm$0&120 \\
BR(\bcl) (8.08$\pm$0.18)\,\%             & $\mp$0&006 & $\mp$0&050 & $\mp$0&105 \\
BR(\bcbarl) (1.62$\pm$0.44)\,\%          & $\pm$0&011 & $\pm$0&174 & $\pm$0&014 \\
BR(\btaul) (0.419$\pm$0.05)\,\%          & $\mp$0&001 & $\pm$0&028 & $\pm$0&007 \\
BR(\cl) (9.77$\pm$0.32)\,\%              & $\pm$0&028 & $\mp$0&151 & $\pm$0&008 \\
$\Rb$ (21.647$\pm 0.072$)\,\%            & $\mp$0&003 & $\pm$0&009 &      0&000 \\
$\Rc$ (16.83$\pm 0.47$)\,\%              & $\pm$0&019 & $\mp$0&099 & $\pm$0&002 \\
{\bf Total branching ratios}             &      0&041 &      0&277 &      0&161 \\
\hline
e ID efficiency $\pm$ 4.1\,\%            & $\mp$0&005 & $\mp$0&007 & $\pm$0&004 \\
$\mu$ ID efficiency $\pm$ 3.0\,\%        & $\mp$0&002 & $\mp$0&021 & $\pm$0&022 \\
Conversions $\pm$15\,\%                  & $\pm$0&001 & $\pm$0&012 & $\mp$0&007 \\
Muon background (K) $\pm$30\,\%          & $\pm$0&011 & $\pm$0&023 & $\mp$0&034 \\
Muon background ($\pi$) $\pm$5\,\%       & $\pm$0&004 & $\pm$0&026 & $\mp$0&015 \\
Muon background (other) $\pm$100\,\%     & $\pm$0&009 & $\pm$0&017 & $\mp$0&058 \\
Background asymmetry u,d,s events        & $\pm$0&002 & $\pm$0&102 & $\pm$0&002 \\
Background flavour separation            & $\pm$0&020 & $\pm$0&020 &   0&000 \\
{\bf Total background effects}           &      0&026 &      0&114 &      0&073 \\
\hline
Correction of $\bar \chi$ for \bcl\      & $ - $0&008 & $ - $0&069 & $ - $0&068 \\
\cthr\ dependence (fractions)            & $ - $0&018 & $ + $0&001 & $ + $0&000 \\
Tracking resolution                      & $ - $0&025 & $ - $0&046 & $ - $0&016 \\
Flavour separation variables             & $\pm$0&067 & $\pm$0&163 & $\pm$0&099 \\
Time dependent mixing                    & $ + $0&032 & $ + $0&098 & $ - $0&091 \\
Monte Carlo statistics                   & $\pm$0&050 & $\pm$0&041 & $\pm$0&070 \\
Charge reconstruction                    & $\pm$0&015 & $\pm$0&004 &      0&000 \\
LEP centre-of-mass energy                & $\pm$0&010 & $\pm$0&027 &      0&000 \\
QCD correction                           & $\pm$0&061 & $\pm$0&036 &      0&000 \\
{\bf Total other systematics}            &      0&114 &      0&216 &      0&167 \\
\end{tabular}
\end{center}
\caption{Results and breakdown of 
systematic uncertainties for the on-peak forward-backward
asymmetries and the mixing parameter \mix.}
\label{tab:systpeak}
\end{table}

\begin{table}[p]
\begin{center}
\begin{tabular}{l|r@{.}l|r@{.}l|r@{.}l|r@{.}l}
& \mco{4}{|c|}{\bf Peak--2} 
& \mco{4}{|c}{\bf Peak+2} \\
& \mco{2}{|c|}{\Afbb (\%)} 
& \mco{2}{|c|}{\Afbc (\%)} 
& \mco{2}{|c|}{\Afbb (\%)} 
& \mco{2}{|c}{\Afbc (\%)} \\ \hline
Fitted Value & 4&70 & -6&83 & 10&31 & 14&59 \\
Statistical error & $\pm$1&80 & $\pm$2&52 & $\pm$1&50 & $\pm$2&04 \\
Systematic error & $\pm$0&098 & $\pm$0&928 & $\pm$0&224 & $\pm$0&837 \\ \hline  
\mco{9}{c}{ } \\
\mco{9}{c}{\bf Sources of systematic errors} \\ \hline
\bl\ semileptonic decay model            & $\pm$0&031 & $\mp$0&036 & $\mp$0&020 & $\mp$0&101 \\
\cl\ semileptonic decay model            & $\mp$0&019 & $\pm$0&070 & $\pm$0&106 & $\mp$0&108 \\
$\mxeb$ $\pm$0.008                       & $\mp$0&008 & $\mp$0&037 & $\mp$0&019 & $\mp$0&034 \\
$\mxec$ $\pm$0.008                       & $\mp$0&013 & $\pm$0&009 & $\pm$0&090 & $\mp$0&051 \\
{\bf Total models}                       &      0&039 &      0&088 &      0&142 &      0&160 \\
\hline
BR(\bl) (10.65 $\pm$0.23)\,\%            & $\mp$0&001 & $\mp$0&034 & $\mp$0&012 & $\pm$0&198 \\
BR(\bcl) (8.08$\pm$0.18)\,\%             & $\mp$0&003 & $\mp$0&047 & $\mp$0&024 & $\mp$0&036 \\
BR(\bcbarl) (1.62$\pm$0.44)\,\%          & $\mp$0&023 & $\pm$0&024 & $\pm$0&055 & $\pm$0&250 \\
BR(\btaul) (0.419$\pm$0.05)\,\%          & $\mp$0&003 & $\mp$0&002 & $\mp$0&002 & $\pm$0&044 \\
BR(\cl) (9.77$\pm$0.32)\,\%              & $\pm$0&009 & $\pm$0&161 & $\pm$0&033 & $\mp$0&381 \\
$\Rb$ (21.647$\pm 0.072$)\,\%            & $\mp$0&001 & $\mp$0&010 & $\mp$0&004 & $\pm$0&023 \\
$\Rc$ (16.83$\pm 0.47$)\,\%              & $\pm$0&006 & $\pm$0&106 & $\pm$0&023 & $\mp$0&250 \\
{\bf Total branching ratios}             &      0&026 &      0&203 &      0&074 &      0&560 \\
\hline
e ID efficiency $\pm$ 4.1\,\%            & $\mp$0&001 & $\pm$0&005 & $\mp$0&007 & $\mp$0&019 \\
$\mu$ ID efficiency $\pm$ 3.0\,\%        & $\mp$0&002 & $\pm$0&030 & $\mp$0&002 & $\mp$0&049 \\
Conversions $\pm$15\,\%                  & $\pm$0&003 & $\mp$0&002 & $\pm$0&003 & $\pm$0&025 \\
Muon background (K) $\pm$30\,\%          & $\pm$0&003 & $\mp$0&032 & $\pm$0&010 & $\pm$0&073 \\
Muon background ($\pi$) $\pm$5\,\%       & $\pm$0&003 & $\mp$0&036 & $\pm$0&005 & $\pm$0&056 \\
Muon background (other) $\pm$100\,\%     & $\pm$0&002 & $\mp$0&044 & $\pm$0&009 & $\pm$0&062 \\
Background asymmetry u,d,s events        & $\pm$0&018 & $\pm$0&832 & $\pm$0&006 & $\pm$0&302 \\
Background flavour separation            & $\pm$0&020 & $\pm$0&025 & $\pm$0&010 & $\pm$0&065 \\
{\bf Total background effects}           &      0&028 &      0&836 &      0&021 &      0&333 \\
\hline
Correction of $\bar \chi$ for \bcl\      & $ - $0&000 & $ - $0&033 & $ - $0&027 & $ - $0&079 \\
\cthr\ dependence (fractions)            & $ + $0&005 & $ - $0&024 & $ - $0&030 & $ + $0&001 \\
Tracking resolution                      & $ - $0&011 & $ + $0&261 & $ - $0&002 & $ - $0&215 \\
Flavour separation variables             & $\pm$0&029 & $\pm$0&172 & $\pm$0&097 & $\pm$0&403 \\
Time dependent mixing                    & $ - $0&037 & $+$  0&036 & $ + $0&069 & $ + $0&112 \\
Monte Carlo statistics                   & $\pm$0&058 & $\pm$0&102 & $\pm$0&064 & $\pm$0&118 \\
Charge reconstruction                             & $\pm$0&007 & $\pm$0&005 & $\pm$0&016 & $\pm$0&010 \\
LEP centre-of-mass energy                & $\pm$0&007 & $\pm$0&018 & $\pm$0&003 & $\pm$0&007 \\
QCD correction                           & $\pm$0&030 & $\pm$0&043 & $\pm$0&062 & $\pm$0&092 \\
{\bf Total other systematics}            &      0&082 &      0&337 &      0&155 &      0&500 \\
\end{tabular}
\end{center}
\caption{Results and breakdown of systematic uncertainties for the off-peak 
forward-backward asymmetries.}
\label{tab:systoff}
\end{table}

\subsection{Modelling of b and c production and decay}

\begin{description}

\item[Semileptonic decay models:]
A correct description of the lepton momentum spectra in the rest frame of 
decaying b and c hadrons is crucial for the flavour separation.
The semileptonic decays of heavy hadrons were described by
the free-quark model of Altarelli \etal\ (ACCMM)~\cite{ref:accmm},
with its two free parameters fixed by fits to 
CLEO data~\cite{ref:cleo} for \bl\ decays
and the combined measurements of  DELCO~\cite{ref:delco}
and MARK III~\cite{ref:markIII} for \cl\ decays~\cite{lephfew}.
For cascade decays \bcl\ and \bcbarl, the D momentum spectrum measured by
CLEO~\cite{ref:cleocasc} is combined with  the \cl\ model to generate
the lepton momentum distribution.  Uncertainties
in the CLEO D momentum spectrum are negligible compared
with the uncertainties in the \bl\ and \cl\ models.

For \bl\ decays, the systematic uncertainties were assessed by
reweighting the rest frame momentum spectrum
according to the form-factor model of
Isgur \etal~\cite{ref:isgw}, which has no free
parameters (ISGW). This spectrum is harder than the central
ACCMM model.
As an alternative, the same model
was used with the fraction of 
\Dss\ mesons produced in b decays increased from 11\,\% to 32\,\%
to describe better CLEO data, denoted ISGW**. This gives a 
softer spectrum than the default model.
Although the 
weights to go from model to model
were evaluated using only \Bzero\ and \Bplus\ meson decays,
these weights were then applied to all b-hadron 
decays.
Variations in the \cl\ decay model were assessed by varying the 
free parameters of the ACCMM model to give harder or softer
decay models, constrained by the DELCO and MARK III data. 
For both \bl\ and \cl\ decays,
the sign of the error assigned indicates the
variation observed with the harder alternative spectrum.

\item[Heavy quark fragmentation:] The lepton momentum spectra depend on
the energy distributions of the heavy hadrons produced in 
\bb\ and \cc\ events.
The Monte Carlo events were reweighted so as to vary the average scaled energy
\meanxe\ of weakly decaying b hadrons in the range $\meanxe=0.702\pm
0.008$ and of c hadrons in the range $\meanxe=0.484\pm
0.008$~\cite{lephfew}.
The fragmentation functions of Peterson~\etal, Collins and Spiller, 
Kartvelishvili~\etal\ 
and the Lund group \cite{fpeter,fragall} were each used as models to 
determine the event weights, and the largest observed variations
were assigned as the systematic errors. For b fragmentation,
the largest shifts were observed with the function of Collins
and Spiller, and for c fragmentation with the function
of Peterson~\etal

\end{description}

\subsection{Branching ratios and partial widths}

\begin{description}
\item[Semileptonic branching ratios:]
The values and uncertainties used for the branching ratios of semileptonic 
b and c hadron decays are listed in 
Tables~\ref{tab:systpeak} and~\ref{tab:systoff} \cite{PDG,lephfew}.

\item[Partial widths of the Z:]
The fractions of hadronic \Zzero\ decays to \bb\ and \cc\ were 
varied according to the uncertainties in 
the LEP average values~\cite{PDG}:
$\Rb = 0.21647 \pm 0.00072 $ and $\Rc = 0.1683\pm 0.0047$. In each case,
the fraction of \Zzero\ hadronic decays to light quarks was adjusted to
compensate the variation.
\end{description}

\subsection{Background uncertainties}\label{s:systbg}

\begin{description}
\item[Lepton identification:]
The analysis is only weakly sensitive to the lepton identification efficiency,
since this mainly affects the ratio of prompt to background leptons, which 
is effectively determined from the data via the fit of the background level.
However, a small component of the background is composed of genuine leptons,
so the lepton identification efficiencies were varied by $\pm 4.1$\,\% for 
electrons and $\pm 3.0$\,\% for muons \cite{opalrb}.

\item[Background composition:]
The overall background fraction is a fitted parameter. However,
the background shape as a function of \NETb, \NETc\ and
$|y|$ may be different for different contributions to the 
background. The rate
of untagged photon conversions was varied by $\pm 15\,\%$ \cite{opalrb}.
In addition, the
$|y|$ distribution of conversions was varied according to the
difference in the rate of identified conversions
observed between data and Monte Carlo simulation.
The contributions to the muon background 
from $\pi$, K and other sources
were varied in turn by 5\,\%, 30\,\% and 100\,\%. These
uncertainties were evaluated using control samples of 
$\mathrm{K^0_s} \rightarrow \pi^+ \pi^-$, three-prong tau decays and
tracks with an enhanced kaon fraction
selected using \dEdx\ cuts \cite{opalrblept}. The variation of the
source fractions, including backgrounds,
as a function of \cthr\ is discussed in Section~\ref{sec:other} below.

\item[Background asymmetries:]
The uncertainties in the background asymmetry were evaluated by 
varying the light primary quark forward-backward asymmetries by the 
uncertainties listed in Table~\ref{tab:bkgasyinput}. Note
that the up and down asymmetries measured by OPAL are +91\,\%
correlated~\cite{opallight}.
They were therefore varied at the same time, and the
resulting
uncertainty combined in quadrature with the uncertainty in the 
strange asymmetry.
The \bb\ and \cc\ asymmetries are treated self-consistently 
in the fit, and the uncertainty in these quantities is therefore
automatically reflected in a very small contribution to the
statistical uncertainty.

\item[Background flavour separation:]
The dependence of the background asymmetry on \NETb\ and \NETc\ is neglected
in the fit (see Equation~\ref{e:bgeq}) but any residual bias is removed
by the bias correction as discussed in Section~\ref{s:res}, providing the
Monte Carlo gives an adequate description of the variations. To quantify
the size of any mismodelling, an additional fit was performed using a 
binned background asymmetry without changing the bias correction, and half the 
difference between this and the standard fit result was assigned as a 
systematic error.

\end{description}

\subsection{Other uncertainties}\label{sec:other}

\begin{description}

\item[Correction factor for mixing in cascade decays:]
The difference between using the standard value of $\oeta=1.083$ and
setting it equal to $1.000$ was assigned as a conservative estimate
of the uncertainty.

\item[Dependence of source fractions on {\boldmath $\cos\theta_{\rm T}$}:]
The description of the source fractions as a function of
\cthr\ was checked by selecting \bl, \cl\ and background
enriched regions of the NETb--NETc plane, and comparing the
\cthr\ distributions of data and Monte Carlo simulation.
No systematic trend was seen for the \cl\ enriched sample. The ratio 
for background varied by up to 10\,\% in the endcap region, with smaller
deviations for the \bl\ enriched sample. A systematic
uncertainty was therefore assessed by reweighting the Monte
Carlo background according to the data/Monte Carlo ratio
observed for the background enriched sample, and repeating
the fit. The change in the background sample was compensated
by an increase in the prompt leptons, so this procedure also
accounted for any possible discrepancy in the \bl\ sample.

\item[Tracking resolution:]
A global 10\,\% degradation in the tracking resolution was applied to
the Monte Carlo sample, separately in the $r$-$\phi$ and $r$-$z$ planes,
as discussed in \cite{opalrb,ref:ojet}. 
The fit was repeated with each of these 
 modified samples, and the sum in quadrature of the two shifts in the 
results was taken as a systematic uncertainty.

\item[Flavour separation variables:]
Possible uncertainties due to mismodelling of the neural network
input variables were taken into account by reweighting the Monte Carlo
events by the ratio between the data and Monte Carlo distributions
of each variable in turn, and repeating the fit. The sum in 
quadrature of the shifts in the results was assigned as a 
systematic uncertainty, dominated by the result of reweighting the lepton
$p$ and $p_t$ distributions for the asymmetries, and 
the input parameter significance for $\mix$.
The neural network output distributions
were also reweighted, to account for the discrepancies between 
data and Monte Carlo visible in Figure~\ref{fig:netdist}, but the 
resulting changes in the results were much smaller than those seen 
when reweighting the inputs, and no additional systematic error
was assigned.

\item[Time dependent mixing:]
The mixing parameter $\mix$ reflects the fraction of mixed events in
an unbiased sample of all b hadrons. However, the time dependent
oscillations for $\Bzero$ mesons are sufficiently slow that the
lifetime variables used in NETc might change the effective value of
$\mix$ as a function of NETc. To evaluate the impact of such an
effect, the value of $\mix$ was evaluated for all $\bb$ Monte Carlo
events with a lepton candidate, in bins of NETc, and for the entire
$\bb$ sample. The fitted value of $\mix$ in the likelihood function
was then scaled by the ratio of Monte Carlo values of $\mix$ in the
NETc bin and $\mix$ in all $\bb$ events. The shift in the fit results
with this rescaling was taken as a systematic uncertainty.

\item[Monte Carlo statistics:]
The fit to the data was repeated 1000 times, with the Monte Carlo reference
distributions being randomly varied according to the statistical
uncertainty in each bin of NETb, NETc and $|y|$.
The RMS variation in the results was added in quadrature with the 
Monte Carlo statistical uncertainty on the bias corrections discussed
in Section~\ref{s:res} to give the total
uncertainty due to limited Monte Carlo statistics.

\item[Lepton charge reconstruction and asymmetry:]
Monte Carlo simulation predicts that about 0.03\,\% of electrons and 0.3\,\% of
muons are reconstructed in the tracking chamber with the wrong charge
(electron tracks are subject to tighter quality requirements because of
the stronger \dEdx\ cuts). This
effect is accounted for in the fit, and the full size of the 
correction is taken as a systematic error.
Studies in \cite{opalbosc}
show a possible difference of around $10^{-3}$ between the lepton 
identification
efficiencies for positive and negative tracks. When combined with the small
difference (2\,\%) in numbers of reconstructed leptons in positive and 
negative hemispheres, this leads to negligible  asymmetry biases of 
around $2\times 10^{-5}$.

\item[LEP centre-of-mass energy:]
The LEP centre of mass energy is known to a precision varying between 3.5 and
20\,MeV depending on the energy point and year \cite{lepebeam}.
Taking year-to-year 
correlations into account, and assuming the Standard Model dependencies of
the asymmetries on $\sqrt{s}$, this leads to the uncertainties given
in Tables~\ref{tab:systpeak} and~\ref{tab:systoff} on the asymmetries
at the quoted values of $\sqrt{s}$.

\item[QCD and thrust axis correction:]
The corrections applied to the raw fitted asymmetries are known to precisions
of 0.80\,\% for b quarks and 0.89\,\% for c quarks, as discussed in 
Section~\ref{s:fitres}. The theoretical part of this uncertainty is
assigned as the error on the QCD correction, whilst the 
Monte Carlo statistical uncertainty is accounted for separately as 
discussed above.

\end{description}

\subsection{Consistency checks}

The analysis was performed separately for data from 1990 and 1991,
for each year separately from 1992 to 1995, and for 
data from 1996 to 2000.
The results were consistent among themselves
and with the standard result from fitting all data simultaneously.
Taking into account statistical errors only,
the  $\chi^2$/dof value for the six \Afbb\ measurements
to be consistent with the same value was 5.9/5. The equivalent
$\chi^2$/dof values 
for \Afbc\ and \mix\ were 2.4/5 and 5.3/5. 
Separate fits for electron and muon events
were also made and showed similar consistency.
The number of bins in $y$, NETb and NETc were 
varied, and again consistent results were obtained. To verify the 
correctness of the fit method itself, it was tested on large samples
of test Monte Carlo samples generated with various b and c asymmetries
and with source fractions and input NETb and NETc distributions generated
to correspond to those in the full simulation. In these tests,
the fit was found to be unbiased and to return correct statistical 
errors on the fitted parameters.

\section{Conclusions}

\begin{figure}
\vspace{-20mm}

\setlength{\epsfxsize}{0.95\hsize}
\epsfbox{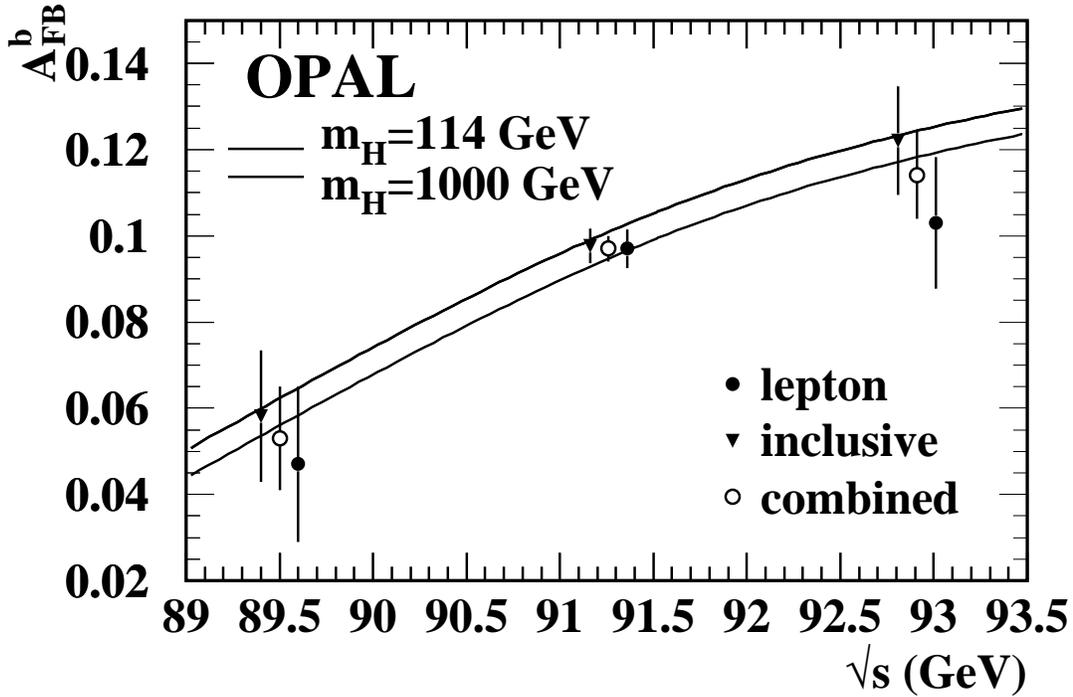}

\vspace{-29mm}

\setlength{\epsfxsize}{0.95\hsize}
\epsfbox{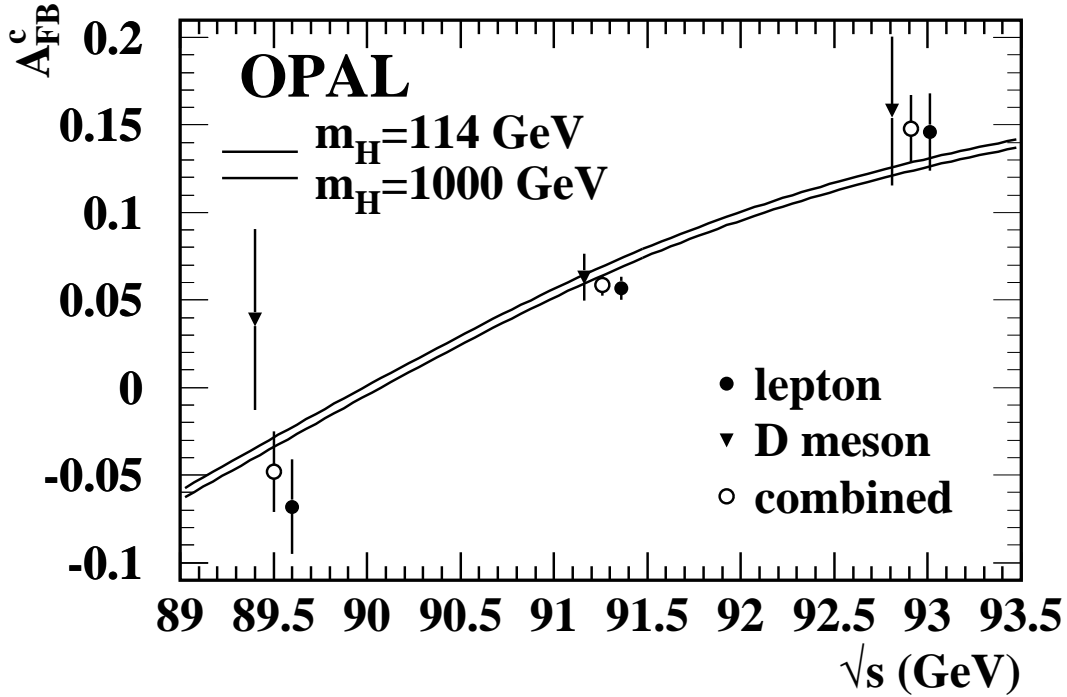}

\caption{\label{fig:rese}The measured b and c quark asymmetries as a function
of centre-of-mass energy, for the lepton, inclusive (b only) \cite{ref:ojet}
and D meson (c only) \cite{ref:odsac}
analyses. The measurement of the b quark asymmetry from D mesons is of
low precision and is not shown. The measurements from different analyses
are slightly displaced on the horizontal axis for clarity.
The combination of the lepton analysis and the other 
analyses is also shown, together with the Standard Model expectation for
Higgs masses between 114 and 1000\,GeV calculated 
using ZFITTER \cite{zfitter}. }
\end{figure}

The forward-backward asymmetries of  \eebb\ and \eecc\ events have been
measured at three energy points around the Z peak, using electrons and muons 
produced in semileptonic decays of bottom and charm hadrons. The results,
corrected to the primary quark level, are:
\begin{center}
\begin{tabular}{@{\Afbb = (}r@{.}l@{$\,\pm\,$}r@{.}l@{$\,\pm\,$}r@{.}l@{)\,\%\ \ \ 
                  \Afbc = (}r@{.}l@{$\,\pm\,$}r@{.}l@{$\,\pm\,$}r@{.}l@{)\,\%\ \ \ 
                  at $ \langle \protect\sqrt s \rangle = $ }r@{.}l@{}}
4&7 & 1&8 & 0&1 & $-$6&8 & 2&5 & 0&9 & 89&51 GeV,\\
9&72 & 0&42 & 0&15 & 5&68 & 0&54 & 0&39 & 91&25 GeV,\\
10&3 & 1&5 & 0&2 & 14&6 & 2&0 & 0&8 & 92&95 GeV. \\
\end{tabular}
\end{center}
For the average B mixing parameter, a value of:
$$
\mix = (13.12 \pm 0.49 \pm 0.42)\,\%
$$
is obtained. In each case the first error
is statistical and the second systematic. The asymmetry 
results are shown as a function of $\sqrt{s}$ in Figure~\ref{fig:rese}.

Using the ZFITTER prediction
for the dependence of \Afbb\ and \Afbc\ on $\sqrt{s}$, the measurements are
shifted to $m_{\rm Z}$ (91.19\,GeV), averaged and corrected for initial
state radiation, $\gamma$ exchange, $\gamma-Z$ interference and quark
mass effects. The results for the pole asymmetries are:
\begin{eqnarray*}
\Afbzb & = & (9.81 \pm 0.40 \pm 0.15)\,\% \, , \\
\Afbzc & = & (6.29 \pm 0.52 \pm 0.38)\,\% \, .
\end{eqnarray*}
These results are consistent with those of \Afbb\ derived from inclusive jet 
charge measurements~\cite{ref:ojet} and those of \Afbb\ and
\Afbc\ derived from fully reconstructed 
charm hadrons~\cite{ref:odsac}.
Whilst the statistical correlations between the asymmetries 
measured 
here and those from reconstructed charm hadrons are negligible, the jet charges
of some of the events involving \bl\ decays are also used to measure \Afbb\
in \cite{ref:ojet}. This results in a statistical correlation of $0.11\pm0.12$
between the two measurements, evaluated using large samples of Monte Carlo
simulated events. 
Taking both statistical and systematic correlations into
account, and making very small corrections to the inclusive and charm meson
asymmetries to use the same nominal centre-of-mass energies as in this analysis,
the final OPAL results for the b and c quark asymmetries around
the Z resonance are:
\begin{center}
\begin{tabular}{@{\Afbb = (}r@{.}l@{$\,\pm\,$}r@{.}l@{$\,\pm\,$}r@{.}l@{)\,\%\ \ \ 
                  \Afbc = (}r@{.}l@{$\,\pm\,$}r@{.}l@{$\,\pm\,$}r@{.}l@{)\,\%\ \ \ 
                  at $ \langle \protect\sqrt s \rangle = $ }r@{.}l@{}}
5&3 & 1&2 & 0&1 & $-$4&8 & 2&2 & 0&7 & 89&51 GeV, \\
9&69 & 0&29 & 0&13 & 5&83 & 0&49 & 0&32 & 91&25 GeV, \\
11&4 & 1&0 & 0&2 & 14&7 & 1&8 & 0&7 & 92&95 GeV, \\
\end{tabular}
\end{center}
and the pole asymmetries are measured to be:
\begin{eqnarray*}
\Afbzb & = & (9.89 \pm 0.27 \pm 0.13) \,\%\, , \\
\Afbzc & = & (6.57 \pm 0.47 \pm 0.32) \,\%\, .
\end{eqnarray*}
In all cases, the first error is statistical and the second systematic.
The $\chi^2$/dof of the fit to all OPAL asymmetry measurements to give
the two pole asymmetries is 11.2/13. (There are six 
asymmetry measurements from 
the lepton tag analysis, three from the inclusive analysis, and 
six from the charm meson analysis, which consists of a simultaneous
fit to the b and c quark asymmetries.) 
Taking into account statistical
and systematic uncertainties, these average pole asymmetries have a 
$+15$\,\% correlation.
Within the framework of the Standard Model, the measurements of \Afbzb\
and \Afbzc\ taken together can be interpreted as a measurement of the 
effective weak mixing angle for leptons of
\[
\swsqeffl = 0.23238 \pm 0.00052 \, .
\]
As can be seen from Figure~\ref{fig:rese}, this result favours large values of
the Higgs mass, in agreement with other measurements of heavy flavour 
asymmetries and in contrast to measurements of the leptonic forward-backward
and left-right asymmetries \cite{ref:s02ew}.

\section*{Acknowledgements}

We particularly wish to thank the SL Division for the efficient operation
of the LEP accelerator at all energies
 and for their close cooperation with
our experimental group.  In addition to the support staff at our own
institutions we are pleased to acknowledge the  \\
Department of Energy, USA, \\
National Science Foundation, USA, \\
Particle Physics and Astronomy Research Council, UK, \\
Natural Sciences and Engineering Research Council, Canada, \\
Israel Science Foundation, administered by the Israel
Academy of Science and Humanities, \\
Benoziyo Center for High Energy Physics,\\
Japanese Ministry of Education, Culture, Sports, Science and
Technology (MEXT) and a grant under the MEXT International
Science Research Program,\\
Japanese Society for the Promotion of Science (JSPS),\\
German Israeli Bi-national Science Foundation (GIF), \\
Bundesministerium f\"ur Bildung und Forschung, Germany, \\
National Research Council of Canada, \\
Hungarian Foundation for Scientific Research, OTKA T-038240, 
and T-042864,\\
The NWO/NATO Fund for Scientific Research, the Netherlands.\\
  

\clearpage


\begin{thebibliography}{99}

\bibitem{ref:s02ew}
The LEP Collaborations,
 ALEPH, DELPHI, L3, OPAL,
    the LEP Electroweak Working Group,
and the SLD Heavy Flavour Group,
`A Combination of Preliminary 
               Electroweak Measurements and 
            Constraints on the Standard Model',
       CERN-EP/2002-091,
       hep-ex/0212036,
 17th December 2002,
and references therein.

\bibitem{PDG}
Particle Data Group, K. Hagiwara \etal,
\PRD{66}{2002}{010001}. \\
The review includes an article by O.~Schneider on 
$\bob$ mixing.
The update using published results for the 2003 online release
({\tt http://pdg.lbl.gov/2003/contents\_listings.html}) 
was used here for the LEP average values of \Rb\ and \Rc.



\bibitem{ref:olasy}
OPAL Collaboration, G.~Alexander \etal, \ZPC{70}{1996}{357}.

\bibitem{ref:alasy}
ALEPH Collaboration, A. Heister \etal, \EPJ{24}{2002}{177}.
\bibitem{ref:dlasy}
DELPHI Collaboration, P.~Abreu \etal, \ZPC{65}{1995}{569}.

\bibitem{ref:llasy}
L3 Collaboration, O.~Adriani \etal, \PLB{292}{1992}{454};\\
L3 Collaboration, M. Acciarri \etal, \PLB{448}{1999}{152}.

\bibitem{ref:ajet}
ALEPH Collaboration, A.~Heister \etal, \EPJ{22}{2001}{201}.

\bibitem{ref:djasy}
DELPHI Collaboration, P.~Abreu \etal, \EPJ{9}{1999}{367}.

\bibitem{ref:ljet}
L3 Collaboration, M. Acciarri \etal, \PLB{439}{1998}{225}.

\bibitem{ref:ojet}
OPAL Collaboration, G. Abbiendi \etal, \PLB{546}{2002}{29}.

\bibitem{ref:adsac}
ALEPH Collaboration, R.~Barate \etal, \PLB{434}{1998}{415}.

\bibitem{ref:ddasy}
DELPHI Collaboration, P.~Abreu \etal, \EPJ{10}{1999}{219}.

\bibitem{ref:odsac}
OPAL Collaboration, G.~Alexander \etal, \ZPC{73}{1996}{379}.

\bibitem{opaldet}
OPAL Collaboration, K.~Ahmet~\etal, \NIM{A305}{1991}{275};\\
P.P.~Allport~\etal, \NIM{A324}{1993}{34}.

\bibitem{opalsi3d}
P.P.~Allport~\etal, \NIM{A346}{1994}{476}.

\bibitem{opalsilep2}
S.~Anderson~\etal, \NIM{A403}{1998}{326}.

\bibitem{opalrb}
OPAL Collaboration, G.~Abbiendi~\etal, \EPJ{8}{1999}{217}.

\bibitem{jetset}
T.~Sj\"ostrand, \CPC{82}{1994}{74}.

\bibitem{jetsetopt}
OPAL Collaboration, G.~Alexander~\etal, \ZPC{69}{1996}{543}.

\bibitem{fpeter}
C.~Peterson, D.~Schlatter, I.~Schmitt and P.M.~Zerwas, \PRD{27}{1983}{105}.

\bibitem{ref:accmm}
G. Altarelli~\etal, \NPB{208}{1982}{365}.

\bibitem{ref:cleo}
CLEO Collaboration, S. Henderson \etal, \PRD{45}{1992}{2212}.

\bibitem{ref:delco}
DELCO Collaboration, W. Bacino \etal, \PRL{43}{1979}{1073}.

\bibitem{ref:markIII}
MARK III Collaboration, R.M. Baltrusaitis \etal, \PRL{54}{1985}{1976}.

\bibitem{gopal}
J.~Allison~\etal, \NIM{A317}{1992}{47}.

\bibitem{jetnet}
The neural networks were trained using JETNET 3: \\
C.~Peterson, T.~R\"ognvaldsson and L.~L\"onnblad, \CPC{81}{1994}{185}.

\bibitem{jetcone}
OPAL Collaboration, R.~Akers~\etal, \ZPC{63}{1994}{197}.

\bibitem{zfitter}
D.~Bardin~\etal, `ZFITTER: An analytical program for fermion pair production
in $\rm e^+e^-$ annihilation', CERN-TH 6443/92, hep-ph/9412201 (1992);\\
D.~Bardin~\etal, \CPC{133}{2001}{229};\\
The input parameters used were $m_{\rm Z}=91.1875$\,GeV, $m_{\rm top}=175$\,GeV
and $m_{\rm Higgs}=150$\,GeV.
 
\bibitem{delphistrange}
DELPHI Collaboration, P.~Abreu \etal, \EPJ{14}{2000}{613}.

\bibitem{opallight}
OPAL Collaboration, K.~Ackerstaff \etal, \ZPC{76}{1997}{387}.

\bibitem{qcdcor}
LEP heavy flavour working group, D.~Abbaneo~\etal, \EPJ{4}{1998}{185}.

\bibitem{qcdcalc}
G.~Altarelli and B.~Lampe, \NPB{391}{1993}{3};\\
V.~Ravindran and W.L.~van Neerven, \PLB{445}{1998}{206};\\
S.~Catani and M.H.~Seymour, \JHP{9907}{1999}{023}.

\bibitem{lephfew}
The LEP collaborations, ALEPH, DELPHI, L3 and OPAL,
\NIM{A378}{1996}{101}.\\
Updated averages are described in `Final input parameters for the LEP/SLD
heavy flavour analyses', LEPHF/2001-01
(see {\tt http://www.cern.ch/LEPEWWG/heavy/} ).

\bibitem{ref:isgw}
    N. Isgur, D. Scora, B. Grinstein and M. Wise, \PRD{39}{1989}{799}.

\bibitem{ref:cleocasc}
CLEO Collaboration, D. Bortoletto \etal, \PRD{45}{1992}{21}.

\bibitem{fragall}
V.G.~Kartvelishvili, A.K.~Likhoded and V.A.~Petrov, \PLB{78}{1978}{615};\\
B.~Andersson, G.~Gustafson and B.~S\"oderberg, \ZPC{20}{1983}{317};\\
P.~Collins and T.~Spiller, \JPH{G11}{1985}{1289}.

\bibitem{opalrblept}
OPAL Collaboration, P.~Acton~\etal, \ZPC{58}{1993}{523}.

\bibitem{opalbosc}
OPAL Collaboration, K.~Ackerstaff~\etal, \ZPC{76}{1997}{401}.

\bibitem{lepebeam}
L.~Arnaudon~\etal, \PLB{307}{1993}{187};\\
R.~Assmann~\etal, \EPJ{6}{1999}{187};\\
LEP energy working group, R.~Assmann~\etal, `Evaluation of the LEP 
centre-of-mass energy for data taken in 2000', LEPEWG 01/01, 
{\tt http://lepecal.web.cern.ch/LEPECAL/}.

\end{thebibliography}
\end{document}